\RequirePackage{fix-cm}
\PassOptionsToPackage{table}{xcolor}
\documentclass{svjour3}               

\smartqed 
\usepackage{graphicx}
\usepackage{url}
\usepackage{todonotes}
\usepackage{enumitem}
\usepackage{amsmath}
\usepackage{soul}
\usepackage{circledsteps}
\usepackage{booktabs}
\usepackage{amssymb} 
\usepackage{fontawesome5}
\usepackage{tabularx}
\usepackage{orcidlink}
\usepackage{subcaption}
\usepackage{siunitx}
\usepackage{multirow}

\newcolumntype{P}[1]{>{\centering\arraybackslash}p{#1}}

\begin{document}

\title{Comparison of Static Analysis Architecture Recovery Tools for Microservice Applications}

\titlerunning{Comparison of Static Architecture Recovery Tools for Microservices}    
\author{Simon Schneider~\textsuperscript{\orcidlink{0000-0001-8605-615X}} \and
        Alexander Bakhtin~\textsuperscript{\orcidlink{0000-0003-3513-7253}} \and
        Xiaozhou Li~\textsuperscript{\orcidlink{}} \and
        Jacopo Soldani~\textsuperscript{\orcidlink{}} \and
        Antonio Brogi~\textsuperscript{\orcidlink{}} \and
        Tomas Cerny~\textsuperscript{\orcidlink{}} \and
        Riccardo Scandariato~\textsuperscript{\orcidlink{0000-0003-3591-7671}} \and
        Davide Taibi~\textsuperscript{\orcidlink{}} 
}

\institute{S. Schneider\at
              Hamburg University of Technology \\
              \email{simon.schneider@tuhh.de}
       \and
           A. Bakhtin \at
              University of Oulu \\
              \email{alexander.bakhtin@oulu.fi}
        \and
           X. Li \at
              University of Oulu \\
              \email{xiaozhou.li@oulu.fi}
        \and
           J. Soldani \at
            University of Pisa \\
              \email{jacopo.soldani@unipi.it}
        \and
           A. Brogi \at
              University of Pisa \\
              \email{antonio.brogi@unipi.it}
        \and
           T. Cerny \at
              University of Arizona \\
              \email{tcerny@arizona.edu}
        \and
           R. Scandariato \at
              Hamburg University of Technology \\
              \email{ricardo.scandariato@tuhh.de}
        \and
           D. Taibi \at
              University of Oulu \\
              \email{davide.taibi@oulu.fi}
}

\date{Received: date / Accepted: date}

\maketitle

\begin{abstract}

Architecture recovery tools help software engineers obtain an overview of the structure of their software systems during all phases of the software development life cycle.
This is especially important for microservice applications because 
they consist of multiple interacting microservices, which makes it more challenging to oversee the architecture.
Various tools and techniques for architecture recovery (also called architecture reconstruction) have been presented in academic and gray literature sources, but no overview and comparison of their accuracy exists.

This paper presents the results of a multivocal literature review with the goal of identifying architecture recovery tools for microservice applications and a comparison of the identified tools' architectural recovery accuracy. 
We focused on static tools since they 
can be integrated into fast-paced CI/CD pipelines.
13 such tools were identified from the literature and nine of them could be executed and compared
on their capability of detecting different system characteristics.
The best-performing tool exhibited an overall F1-score of 0.86.
Additionally, the possibility of combining multiple tools to increase the recovery correctness was investigated, yielding a combination of four individual tools that achieves an F1-score of 0.91.

\vspace{2mm}
\noindent
\textbf{Registered report:} The methodology of this study has been peer-reviewed and accepted as a registered report at MSR'24: \url{https://arxiv.org/abs/2403.06941}

\end{abstract}

\section{Introduction}
\label{sec:introduction}

Static analysis tools can support developers with valuable feedback on their work without the need to run and test their systems. 
Tools such as SonarQube~\footnote{sonarsource.com/products/sonarqube/}, PMD~\footnote{pmd.github.io/}, or IntelliJ~\footnote{jetbrains.com/de-de/idea/} are examples of widely popular solutions that perform real-time analyses for different aspects of development.
Architecture recovery tools (also called \textit{architecture reconstruction}) can support developers by showing the implemented system's high-level architectural design, helping them adhere to the intended design and avoid architectural issues such as violations of security rules~\cite{Tukaram22_security_rules}, bad smells~\cite{Ponce2022_MLRSecuritySmellsMSA}, antipatterns~\cite{Taibi20_microservices_antipatterns}, or non-conformances~\cite{Cao24_catma}.
Providing accessibility of the systems' architecture is additionally important for 
microservice systems, consisting of tens, often hundreds, of interacting microservices.

The microservice architecture is an architectural style that often entails a distributed codebase.
It has been increasingly adopted in the last years and continues to gain popularity in software development.
Applications employing the microservice architecture split their business logic into multiple microservices. 
The individual microservices communicate over lightweight communication channels~\cite{Dragoni16_microservices_yesterday_today_tomorrow,Lewis14_microservices}.
The architecture has many benefits for software engineering activities. 
However, the distributed nature of the codebase can pose challenges, since it is more difficult to gain and maintain an overview of the application's architectural design~\cite{DiFrancesco19_architecting_microservices,Soldani2018_MSAPainsGains}.

Architecture recovery tools and techniques can support developers and software architects in this regard by creating a representation of the implemented system.
In the field of program comprehension, it has been shown that such representations foster better and easier analysis, maintainability, and usability, and support software engineers during development (e.g.,~\cite{Arisholm06_impact_uml,Budgen11_uml_slr,Gravino10_empirical_investigation_code_comprehension,Gravino15_code_comprehension_uml,Schneider24_DFDs_empirical_experiment}).
By allowing to assess the adherence of the implemented architecture to the designed one, issues such as architectural drift are also mitigated.

With the growing adoption of the microservice architecture, the need for static analysis tools that specialize in microservice applications rises.
Consequently, various approaches for architecture recovery for microservices have been proposed in the academic literature~\cite{Alshuqayran18,Granchelli17,Kleehaus18,Queval23_extracting_architecture,Soldani21}.
Some authors also provide tools to show the feasibility of their presented approaches.
Further tools can be found in the gray literature on the topic.

Among these tools, those that follow a static approach (as opposed to dynamic or hybrid approaches) are especially attractive for use in modern software engineering practices, where rapid development cycles are the norm.
In such fast-paced scenarios, lightweight techniques are preferred since they can be better plugged into CI/CD (continuous integration/continuous delivery) pipelines without the need for complex analysis environments and without impeding timely processing.

In this paper, we present a study that we conducted to identify and compare static analysis tools for architecture recovery of microservice applications.
A core contribution is the execution of all identified tools on a common dataset and the comparison of their accuracy in performing architecture recovery.
We measure the extraction correctness in terms of precision, recall, and F1-score of extracted application characteristics, compared to a manually created ground truth.

In the context of this work, we refer to a microservice application's architecture as a representation of its architectural design in terms of components and their logical connections.
The microservice architecture offers such a system decomposition by definition, since individual microservices are meant to be self-standing, independently deployable units.
Communication links between them that are necessary to fulfill the system's business logic form the connections between components.

\paragraph{Research Questions}

In pursuing to fulfill the above objectives, this paper addresses the following research questions:

\begin{itemize}
    \item \textbf{RQ1: Which freely available, static analysis tools for architecture recovery of microservice applications exist?} \\
    Various architecture recovery approaches have been proposed in the literature, which are often supported by prototypes implementing the techniques.
    Additionally, gray literature sources give pointers to further tools that fit this scope.
    We curated a list of such tools in a multivocal literature review and evaluated and compared them on a common benchmark.
    
    \item \textbf{RQ2: Which characteristics do the tools extract in addition to the basic architecture (i.e., components and connections between them)?} \\
    This work is mainly concerned with the reconstruction of the analyzed application's basic architecture, i.e., services and connections between them.
    However, many tools extract additional application \emph{characteristics}, such as information about implemented security mechanisms, links to design requirements, or trust boundaries.
    An overview of these additional characteristics helps to identify tools for specific use cases.
    
    \item \textbf{RQ3: Which are the most commonly considered characteristics extracted by the tools?} \\
    Based on the presentation of the characteristics extracted by the identified tools, we analyzed the tools' overlaps and differences in their extraction scopes.
    The results show what the tools mostly focus on and also where gaps lie.
    
    \item \textbf{RQ4: Which is the identified tools' accuracy in architecture recovery?} \\
    To compare the identified tools based on their correctness in architecture recovery, we executed them on a common benchmark and measured their precision, recall, and F1-score concerning the following properties:
        \begin{itemize}
            \item \textbf{RQ4.1: Which is the identified tools' accuracy in detecting the components that form a microservice application?}
            The individual microservices of an application constitute the building blocks that the application's microservice architecture consists of.
            The components are often the easiest characteristics to extract for microservice applications, since deployment technologies such as Docker Compose or Kubernetes ease their detection.
            Nevertheless, this is the foundational step of architecture recovery and was thus evaluated.
            
            \item \textbf{RQ4.2: Which is the identified tools' accuracy in detecting connections between the components forming a microservice application?} \\
            The second  characteristic in the core architecture of microservice applications is the connections between the components, over which requests are made and data is exchanged.
            Such connections can be realized in different ways, for example via direct API calls, asynchronous communication techniques, or implicit invocations by infrastructural components such as the communication for registering services in a service registry.
            Due to this added complexity, the detection of connections is harder than that of the components and is evaluated separately.
            \item \textbf{RQ4.3: Which is the tools' accuracy in detecting the additionally extracted characteristics?} \\
            For those tools that extract characteristics in addition to the basic architecture (see RQ2), we measured their correctness in extracting this extra information.
            We compared tools that extract the same additional characteristics.
        \end{itemize}
    \item \textbf{RQ5: Can combinations of multiple tools outperform the best individual tools?} \\
    Combinations of multiple individual tools could show synergies that improve the results over those observed when executing the tools alone.
    We investigated multiple combinations of tools concerning the possibility of improving the architecture recovery correctness in all three evaluated metrics (precision, recall, and F1-score).

\end{itemize}

This paper provides an overview and comparison of state-of-the-art, freely available, static analysis architecture recovery tools for microservice applications.
The presented findings can be beneficial to both researchers and practitioners. 
Insights on the quality of such tools, as well as on the specific characteristics they extract, are valuable for both groups of stakeholders.
For researchers, an overview of existing tools can prevent the creation of yet another approach for which a similar technique has already been proposed.
Consequently, the results of the study can shed light on the directions for future work and help accelerate research on the topic.
For practitioners, the results of the study can provide a reference for the tools and for comparing their actual capabilities.
Industry adoption of tools and techniques presented in academic literature is notoriously challenging.
Our study is geared towards fostering visibility of tools for microservice architecture recovery and showing their accuracy.
With the comparison based on results observed from executing all tools on the same applications instead of an overview of their approaches, our study fills a gap in the literature that has not been addressed before, to the best of our knowledge.
Especially in academia, where published tools are often prototypes created for the sake of showing the feasibility of a presented approach and where subsequent maintenance is often neglected, such an evaluation is crucial for properly judging the tools' qualities.

The rest of this paper is structured as follows: Section~\ref{sec:methodology} provides an extended and updated description of the methodology presented in the registered report of this work. Section~\ref{sec:lit_rev} describes the MLR and the tools identified with it. Section~\ref{sec:comparison} presents the comparison of the identified tools concerning their extraction correctness and Section~\ref{sec:discussion} discusses these results and presents lessons learned. Section~\ref{sec:limitations} describes limitations of the presented work. Finally, Section~\ref{sec:related_work} presents the related work and Section~\ref{sec:conclusion} concludes the paper. 

\section{Methodology}
\label{sec:methodology}

The methodology of the conducted study has been presented as a registered report at the International Conference on Mining Software Repositories 2024 (MSR'24)~\cite{Schneider24_registered_report}.
We repeat it below in updated form for consistency of this paper. 
Research question RQ5 and the related methodology (presented in Section~\ref{sub:tool_combinations_methodology}) is an extension with respect to the registered report.

\begin{figure*}
    \centering
    \includegraphics[width = \linewidth]{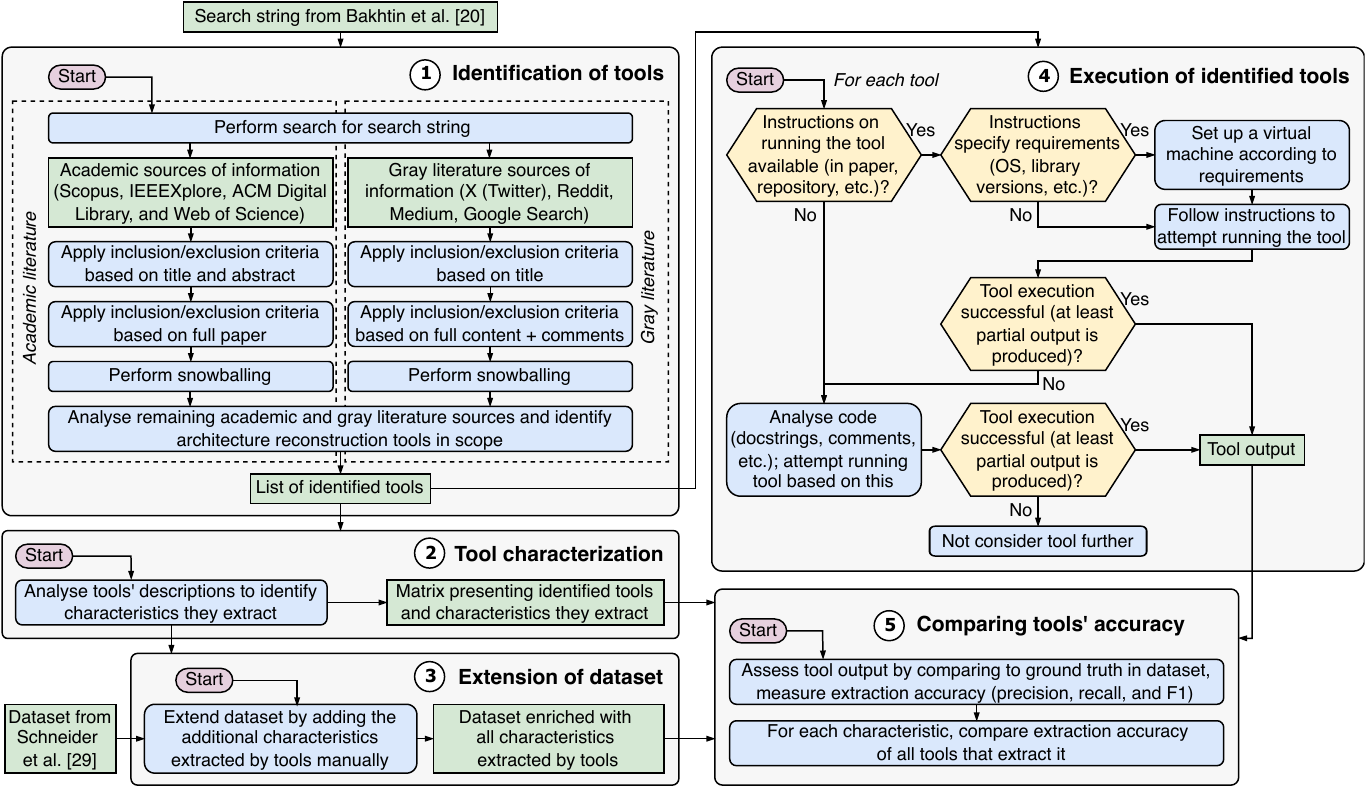}
    \caption{The methodology adopted in this study.}
    \label{fig:methodology}
\end{figure*}

The presented work consisted of two parts, (i) a multivocal literature review to identify static analysis architecture recovery tools for microservice applications, and (ii) a comparison of the identified tools' accuracy in architecture recovery by executing them on a common dataset and evaluating the outputs they produce.
Figure~\ref{fig:methodology} shows the complete methodology structured into five steps.
Each step is further described in the following sections, in the order indicated by the figure.

\subsection{Identification of Tools} 
\label{sub:methodology_tool_selection}

The systematic literature review in this work is a replication of the one presented by Bakhtin et al.~\cite{Bakhtin23_tools_study}.
The search has been repeated to identify tools published after the authors performed their search.
Further, the in- and exclusion criteria of our study have been applied to the tools already identified in the replicated literature review to select those relevant to us.
The original list is not restricted to a specific analysis approaches, while we focus on static analysis tools, i.e., a subset of the original list.
Because of this and since the methodology we applied is adapted from the one of Bakhtin et al.~\cite{Bakhtin23_tools_study}, no relevant tools were missed in this process.
Additionally to academic sources, we adopted the methodology and applied it to gray literature sources as well, thus extending the work into a multivocal literature review~\cite{GAROUSI2019101} (see step~\Circled{1} in Figure~\ref{fig:methodology}).

For the repetition of the literature review of Bahtin et al.\cite{Bakhtin23_tools_study}, we used the same search string (given below) and searched for it in the same four scientific databases (\textsc{Scopus},\footnote{\textsc{Scopus}: \url{https://www.scopus.com}.}, \textsc{IEEEXplore}\footnote{\textsc{IEEEXplore}: \url{https://ieeexplore.ieee.org/}.},  \textsc{ACM Digital Library},\footnote{\textsc{ACM Digital Library}: \url{https://dl.acm.org}.} and \textsc{Web of Science}\footnote{\textsc{Web of Science}:  \url{https://www.webofscience.com/wos/woscc/basic-search}}).
We only considered results published after the search date reported in the paper of the original literature review.

Search string:
\begin{center}
\small
\ttfamily
(Microservice* OR Micro-service* OR "micro-service*") \\
AND Architect* \\
AND (Reconstr* OR Mining OR Reverse engineering \\ OR Recover* OR Extract* OR Discover*) \\
AND (Tool* OR Prototype OR Implementation OR GitHub OR \\ Proof of concept OR POC OR Proof-of-concept)
\end{center}

\noindent
The fourth AND-group of the search string is used for in-text search if the database functionality allows it.

For gray literature sources, we applied the search string to four websites: \textsc{Google Search}\footnote{\textsc{Google Search}: \url{https://google.com}}, \textsc{X (Twitter)}\footnote{\textsc{X (Twitter)}: \url{https://x.com}}, \textsc{Reddit}\footnote{\textsc{Reddit}: \url{https://reddit.com}}, and \textsc{Medium}\footnote{\textsc{Medium}: \url{https://medium.com}}.
These are popular websites for a technical audience and have been used as sources of gray literature resources in the literature~\cite{Moreschini2023TowardEM,PELTONEN2021106571}.
Duplicates were removed during result aggregation.
After compiling the list of initial sources, we applied the following inclusion and exclusion criteria (adapted from Bakhtin et al.~\cite{Bakhtin23_tools_study}), to fit our target scope:

\noindent
\textbf{Inclusion criteria:}
    \begin{enumerate}[leftmargin=*]
        \item Mentions a tool for microservice architecture recovery
        \item A reference to the freely available tool is made or the tool can be found by searching for its name
        \item The tool follows a static or hybrid analysis approach
    \end{enumerate}
\textbf{Exclusion criteria:}
    \begin{enumerate}[leftmargin=*]
        \item Source is not in English
        \item Out of topic -- relevant terms are used in a different context
        \item Source describes different aspects of microservice recovery (not dealing with tools)
        \item Source describes closely related tasks such as monolith to microservice migration
    \end{enumerate}

\noindent
We applied the criteria in two phases, as it is common practice when conducting SLRs~\cite{Kitchenham04_procedures,Kitchenham07_guidelines,Ralph20_empirical_standards}.
For the academic sources, sources were excluded based on reading the title and abstract of the paper in the first phase; in the second phase, the complete paper was examined. 
For the gray literature sources, we considered the title in the first phase and the full content including comments in the second phase.
In each phase, sources were excluded that met any of the formulated exclusion criteria or if it could be decided that the source does not meet all inclusion criteria.
We took a conservative approach to exclusion: sources for which a decision could not be made in the first phase were retained for evaluation in the second phase.

All steps were performed by two authors independently, and disagreements solved via discussion with a third author.
We assessed the authors' agreement with Cohen's kappa coefficient \cite{10.1023/A:1009820201126}. The kappa coefficient considers the observed frequency of agreement relative to the expected probability of agreement, assuming authors decide randomly and independently.

Finally, we performed forward and backward snowballing~\cite{Wohlin} on the academic sources, i.e., we reviewed all references and citations at relevant places of the paper in the same way as described above.
For the gray literature sources, we instead checked whether contained links to other resources refer to tools that are in scope.

We examined the identified sources in detail to identify all presented and mentioned tools for microservice architecture recovery.
Specifically, we looked for any references to source code repositories, web applications, Docker images, or other ways of providing a tool.
In line with the objective of the study, only tools that follow a static analysis approach or hybrid approach where the static part can be run independently were considered.
The first author performed the identification. 
In cases where no tools were found in a source, the second author checked the source as well for confirmation.
The resulting list of static analysis microservice architecture recovery tools serves as the answer to \textbf{RQ1}.

\subsection{Tool Characterization}
\label{subsub:tool_codification}

For all identified tools, we examined the available information (corresponding publication, code repository, tool description, etc.) to identify their general properties (platform, language, static/hybrid approach, output format, etc.), as was done by Bakhtin et al.~\cite{Bakhtin23_tools_study} (step~\Circled{2} in Figure~\ref{fig:methodology}).
Here, those characteristics extracted by the tools that go beyond the basic architecture of the analyzed systems were of particular interest.
The results are presented in a matrix listing all identified tools as well as the characteristics they extract.
This matrix later determined which tools are compared with each other based on each characteristic. 
In addition to guiding the comparison of tools' extraction accuracy, the created matrix is also the basis to answer \textbf{RQ2} and \textbf{RQ3}.

\subsection{Extension of Dataset}
\label{sub:extension_of_dataset}

\begin{table}[t]
    \small
    \centering
	\caption{Microservice applications in the used dataset, their numbers of components (\textbf{Cp}), connections (\textbf{Cn}), and endpoints (\textbf{E}).}
	\label{tbl:evaluation_applications}
    \begin{tabular}{cl|ccc}
        \toprule
        \textbf{App.} & \textbf{GitHub Repository} & \textbf{Cp} & \textbf{Cn} & \textbf{E} \\
        \midrule
        1 & anilallewar/microservices-basics-spring-boot & 12 & 29 & 8 \\
        2 & apssouza22/java-microservice & 15 & 34 & 17 \\
        3 & callistaenterprise/blog-microservices & 17 & 42 & 10 \\
        4 & ewolff/microservice & 7 & 13 & 15 \\
        5 & ewolff/microservice-kafka & 8 & 12 & 11 \\ 
        6 & fernandoabcampos/spring-netflix-oss-microservices/ & 11 & 24 & 17 \\
        7 & georgwittberger/apache-spring-boot-microservice-example & 5 & 6 & 6 \\
        8 & jferrater/tap-and-eat-microservices & 9 & 16 & 11 \\
        9 & koushikkothagal/spring-boot-microservices-workshop & 5 & 6 & 7 \\
        10 & mdeket/spring-cloud-movie-recommendation & 11 & 18 & 17 \\
        11 & mudigal-technologies/microservices-sample & 15 & 34 & 3 \\ 
        12 & piomin/sample-spring-oauth2-microservices & 8 & 13 & 2 \\
        13 & rohitghatol/spring-boot-microservices & 11 & 26 & 8 \\
        14 & shabbirdwd53/springboot-microservice & 9 & 18 & 6 \\
        15 & spring-petclinic/spring-petclinic-microservices & 12 & 28 & 10 \\ 
        16 & sqshq/piggymetrics & 17 & 37 & 10 \\
        17 & yidongnan/spring-cloud-netflix-example & 10 & 29 & 2 \\
        \bottomrule
    \end{tabular}
\end{table}

To compare the identified tools' correctness in architecture recovery under controlled circumstances, they needed to be executed on the same dataset.
We used a dataset of 17 dataflow diagrams (DFDs) of open-source microservice applications~\cite{Schneider23_microsecend} for this purpose.
To the best of our knowledge of the research field, it is the only dataset suited for this purpose. 
We are not aware of any other collection of manually created and verified architectural models in this form.
Table~\ref{tbl:evaluation_applications} lists the applications in the dataset along with their number of components, connections, and endpoints.
They are typical, small- to medium-sized open-source microservice applications written in Java with a focus on the Spring framework.
It has been reported, that Java is the most popular language for developing microservice applications~\cite{JetBrains22} and that Spring is the most used framework for Java microservice applications~\cite{JRebel22}.
The applications in the dataset were selected from sources in the literature as well as popular repositories on GitHub. 
According to the dataset's publication, established design patterns for microservice applications using the Java Spring framework are prevalent in the dataset.

The DFDs in the dataset depict the applications' architecture as well as additional properties. 
Nodes in the DFDs represent components of the applications, i.e., (internal and infrastructural) microservices, databases, and external entities; edges represent connections between any two components.
As such, the nodes and edges were used as ground truth for the basic architecture (i.e., RQ4.1 and RQ4.2 are answered based on the tools' accuracy in extracting these characteristics).

To answer RQ4.3, an extension of the dataset was needed.
As will be presented later in Section~\ref{sub:extracted_characteristics}, the characteristic \textit{endpoints} was the only additional one to components and connections that could be part of the comparison.
In this context, endpoints describe explicitly specified endpoints of components to which RESTful HTTP requests can be made.
The DFDs in the dataset contain extensive annotations that represent security mechanisms, deployment information, and other system properties, including information about endpoints.
However, the information about endpoints was not complete.
Therefore, the DFDs needed to be extended concerning this characteristic to serve as ground truth for evaluating the tools' extraction correctness.
A look into the tools that have endpoints in their extraction scopes and an investigation into possibilities for implementing endpoints revealed, that a number of Java annotations indicate such endpoints.
Specifically, identifying the annotation \texttt{@RequestMapping} and its related, more specific annotations for a single HTTP method (\texttt{@PutMapping}, \texttt{@GetMapping}, and so on) and the annotation \texttt{@RepositoryRestResource} in the code is sufficient to manually create the ground truth required for this work.

Due to this simple and unambiguous nature of the identification of endpoints, a single author searched for the relevant annotations in the source code of all applications in the dataset and extended the DFDs accordingly. 
A second author checked the newly added endpoints for correctness and found no errors.

We identified 63 additional endpoints following this process and added them to the DFDs in the dataset.
As a result, the dataset now contains exhaustive information about endpoints specified with the considered Java annotations.

Since some of this paper's authors are also authors of the initial dataset, we have integrated the described changes directly.
All means of obtaining the dataset have been updated to the new data (the corresponding website~\footnote{https://tuhh-softsec.github.io/microSecEnD/} and GitHub repository~\footnote{https://github.com/tuhh-softsec/microSecEnD} contain the new data, and we added a new version containing the new data to the persistent repository at Zenodo~\footnote{https://zenodo.org/records/7714926}).

\subsection{Execution of Identified Tools}
\label{sec:execution}
To obtain outputs from the identified tools for their evaluation, they were run on the applications in the dataset.
Naturally, the tools needed to be executed successfully to create outputs.
There were some obstacles in terms of reproducibility, i.e., it was not trivial to execute some tools.
To achieve a fair comparison, a methodology for executing them was established (step~\Circled{4} in Figure~\ref{fig:methodology}) that ensures that the effort invested into attempting to run each tool is comparable.
We first attempted to run a tool based on available instructions (documentation, information in the source, etc.), possibly on a virtual machine if specific requirements for the execution environment were mentioned.
Where this was not successful, we analyzed the code for indicators of how to run the tool (code comments, hints by identifiers, error messages during execution, etc.).
If all steps failed, we abandoned the tool and excluded it from the comparison.
As an indicator of a successful execution, we checked whether the tool produced any output in the form of extracted characteristics or status information signaling a completed run.

\subsection{Comparing Tools' Accuracy}
The metrics precision, recall, F1-score, and execution time were used as quantitative measures for comparing the tools.
These are common and objective metrics used for such evaluations.
Although we do not dictate a specific use case for the tools, their ability to perform their core functionality correctly is the most important basis for evaluation.
Ideally, an architecture recovery tool should extract all existing characteristics in its extraction scope and not falsely produce results for more than these.
Precision and recall serve as measures to indicate these two aspects, the F1-score serves as harmonic mean between the two to allow a clear comparison based on a single score.
The relevance of the execution times is more dependent on the intended use case, but is important for most scenarios as well.
Lightweight static analysis tools that show quick execution times lend themselves to being integrated into automated pipelines such as fast-paced CI/CD pipelines.
For example, the output of architecture recovery tools could be used by model-based analysis tools in a deployment pipeline.

To quantify the tools' output, the number of correctly extracted characteristics (true positives, TP), the number of falsely extracted characteristics (false positives, FP), and the number of undetected characteristics (false negatives, FN) were manually counted by comparing the output to the ground truth (step~\Circled{5} in Figure~\ref{fig:methodology}).
Since the tools have different output formats, quantifying the results manually was deemed the safest method for a correct representation.

The process was performed by two authors independently, and disagreements were solved in discussion with a third author.
Precision, recall, and F1-score were calculated from these measures with the following formulas:

\small
\[\text{Precision} = \frac{\text{Correct characteristics}}{\text{Correct characteristics}+\text{False characteristics}}\]

\[\text{Recall} = \frac{\text{Correct characteristics}}{\text{Correct characteristics}+\text{Undetected characteristics}}\]

\[\text{F1-score} = \frac{2 * \text{Correct characteristics}}{2 * \text{Correct characteristics}+\text{False characteristics}+\text{Undetected characteristics}}\]\newline
\normalsize

\textbf{RQ4} is answered based on the above measures.
Specifically, we compared different subsets of the complete list of tools against each other.
The overview of each tool's extracted characteristics (see Section~\ref{subsub:tool_codification}) determined which tools were compared with each other. 
For each characteristic, we compared all tools that are supposed to extract it in extracting this characteristic concerning their accuracy.

The tools' execution times were measured by first creating Bash scripts that execute a tool on all applications on which it can be run.
The scripts consider only the analysis itself, i.e., they exclude the possibly needed cloning of repositories or compilation of the analyzed applications.
The scripts were then run, and the execution times measured using the \texttt{time} terminal command.
Each tool's script was executed ten times and the average execution time calculated.

\subsection{Evaluating Combinations of Tools}
\label{sub:tool_combinations_methodology}

Additionally to evaluating the individual tools, we investigated whether it is possible to improve their results by combining multiple tools.
Tools could show synergies concerning the scope of extracted characteristics and concerning the applied detection techniques.
Therefore, combinations were evaluated by combining the individual tools' results.
The combinations were selected based on the individual results such that the used metrics (precision, recall, and F1-score) are maximized.

In this context, the tools' results are combined with logical ``AND'' and ``OR'' operators on the individual characteristic level, indicated in the following as subscript annotations to the tool combinations' IDs.
For each individual characteristic, combining the results of two or more tools with a logical ``AND'' means, that the characteristic is successfully extracted (TP) if all tools have done so.
Otherwise, it counts as FN.
A FP is counted if all tools falsely detected the same individual characteristic.
Note that there can also be the case that only one of the tools in a combination successfully produced results for an application, in this case, its results count directly for the combination.
For the combinations with a logical ``OR'', it is sufficient if one of the combinations' tools had detected an individual characteristic to count as TP for the combination, and only if none of the tools detected it does it count as FN.
Here, all FPs of the individual tools are considered and those falsely detected by all tools only count as one FP.
Table~\ref{tbl:combination_measures_description} summarizes the above description.
Precision, recall, and F1-measure are calculated in the same way as for the individual tools.

\begin{table}[h!]
\centering
\caption{Definitions of true positives (TP), false negatives (FN), and false positives (FP) for "OR" and "AND" combinations.}
\label{tbl:combination_measures_description}
\begin{tabular}{c|c|p{8cm}}
    \toprule
    \textbf{Combination} & \textbf{Measure} & \textbf{Description} \\
    \midrule
    \multirow{3}{*}{OR} & TP & \textit{Any} tool \textit{correctly} detected the characteristic. \\
                        & FN & \textit{No} tool \textit{correctly} detected the characteristic. \\
                        & FP & \textit{Any} tool \textit{falsely} detected the characteristic. \\
    \midrule
    \multirow{3}{*}{AND} & TP & \textit{All} tools \textit{correctly} detected the characteristic. \\
                         & FN & \textit{Any} tool did \textit{not correctly} detect the characteristic. \\
                         & FP & Al\textit{}l tools \textit{falsely} detected the characteristic. \\
    \bottomrule
\end{tabular}

\end{table}

\section{Tools Identified via a Multivocal Literature Review}
\label{sec:lit_rev}

The multivocal literature review was executed following the methodology described in Section~\ref{sub:methodology_tool_selection} (step \Circled{1} in Figure~\ref{fig:methodology}).
The process and identified tools are presented in the context of formally and informally published sources, with a description of the tools and their characteristics.

\begin{figure}
    \centering
    \includegraphics[width = \linewidth]{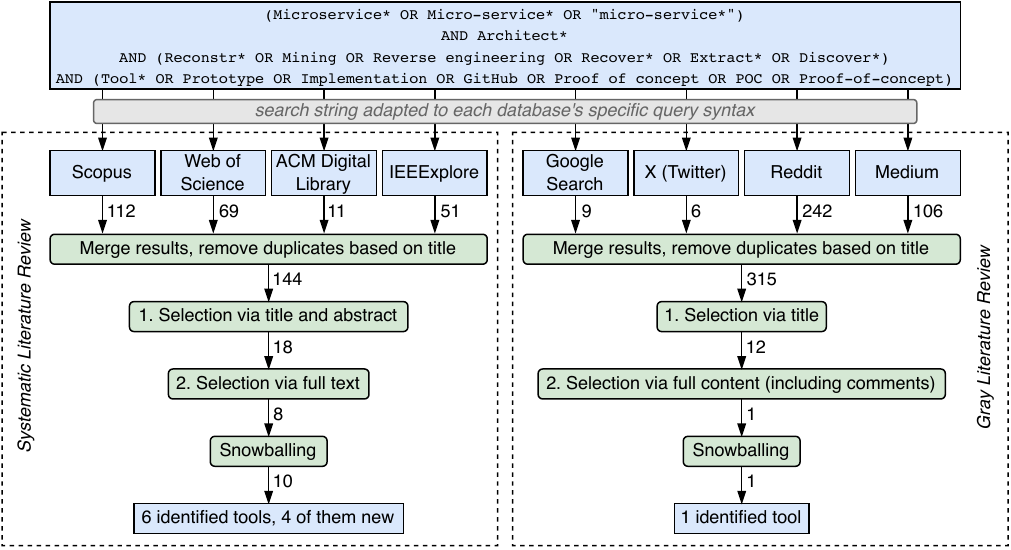}
    \caption{Results of the systematic literature review of the formally published literature (left) and the gray literature review of the informally published literature (right).}
    \label{fig:methodology_results}
\end{figure}

\subsection{Review of Informally Published (Gray Literature) Sources}
\label{sub:gray_lit_review}

Figure~\ref{fig:methodology_results} (right side) visualizes the process of the conducted gray literature review. 
Searching for the formulated search string yielded 315 initial results across the four considered sources.
Most results were found on Reddit (204) and on Medium (106).
The results from Reddit contained mostly posts asking for advice, e.g., with architecting microservice applications, implementing specific technologies, visualizing an architecture, or a reference to a tool that could perform architecture reconstruction. 
The search on Medium resulted in many unrelated posts without any connection to microservices or even computer science.
Those results that were more relevant contained mostly guides on specific implementations or technologies.

As shown in Figure~\ref{fig:methodology_results}, twelve results were retained after applying the exclusion criteria in the first phase, i.e., after reading the results' titles and descriptions.
After reading the complete results in the second phase, all but one result were excluded.
The only result that is in the scope of the study contains a reference to a proprietary tool that meets the inclusion criteria.
The tool is not open-source, but a free use plan exists.
While we would not be able to perform any debugging steps in case of execution problems (see step \Circled{4} in Figure~\ref{fig:methodology}), the tool meets the defined scope of our study, and we thus opted to include it in the further process.

The calculated Cohen's Kappa for interrater agreement showed almost perfect agreement for the first phase ($\kappa$ = 0.84) and perfect agreement for the second phase ($\kappa$ = 1.0) according to Landis and Koch's classification~\cite{Landis77_agreement}.
The reason for such a high agreement was due to the rather unambiguous inclusion decision to be made -- identifying a reference to a tool that is in the scope of our study or concluding that such a reference is not given is a straightforward task when assessing the complete paper.

\subsection{Review of Formally Published (Academic) Literature}
\label{sub:slr}

For the identification of tools in the academic literature, we performed a systematic literature review following the methodology described in Section~\ref{sub:methodology_tool_selection}.
Figure~\ref{fig:methodology_results} (left side) shows the intermediate results of each step.
The reported numbers correspond to results that were published after the search date reported by Bakhtin et al.~\cite{Bakhtin23_tools_study}.
The search in the four formal databases yielded 144 results.
After applying the inclusion/exclusion criteria based on the articles' title and abstract in the first phase, 18 articles remained. 
Applying the criteria on the articles' full content in the second phase retained 8 articles.
We added one source known to us, which we know to contain a tool in the scope of our study (\textit{MicroGraal} \cite{kozak2024micrograal}).
It was not found via the search because it was only recently presented.
A following forwards and backwards snowballing revealed one further candidate.

From the ten sources, six tools could be identified, however, two of those had already been found with the previous study.
Their occurrence in both literature reviews that should have mutually exclusive results due to the considered time periods is explained by pre-prints that had been published before the cut-off date and the final publications being published after.
Thus, four new tools have been identified in the SLR, compared to the previous list of tools.

The Cohen's Kappa coefficient for interrater agreement indicated substantial agreement for both phases ($\kappa$ = 0.77 for both)~\cite{Landis77_agreement}.

\subsection{Identified Tools}
\label{sub:identified_tools}

\begin{table}
    \footnotesize
    \centering
	\caption{Identified static analysis architecture recovery tools. \\ \textbf{Pub.} = Publication; \textbf{Comp.} = considered in comparison of tools, reasons for exclusion are discussed in the presentation of tools below.}
	\label{tbl:tools}
    \begin{tabular}{p{0.6cm}p{3.3cm}p{4.5cm}P{0.6cm}P{0.8cm}}
        \toprule
        \textbf{ID} & \textbf{Name} & \textbf{GitHub Repository} & \textbf{Pub.} & \textbf{Comp.}\\
        \midrule
        AFA & \textit{authz-flow-analysis} & cloudhubs/authz-flow-analysis & \cite{abdelfattah2023towards} & $\square$ \\
        AGG & \textit{Attack Graph Generator} & tum-i4/attack-graph-generator & \cite{Ibrahim19_attack-graph-generator} & $\blacksquare$ \\
        C2D & \textit{Code2DFD} & tuhh-softsec/code2DFD & \cite{Schneider23_code2dfd} & $\blacksquare$ \\
        MDG & \textit{MicroDepGraph} & clowee/MicroDepGraph & \cite{Rahman19_microdepgraph} & $\blacksquare$ \\
        MGR & \textit{MicroGraal} & cloudhubs/graal-prophet-utils & \cite{kozak2024micrograal} & $\blacksquare$ \\
        MMI & \textit{microMiner} & di-unipi-socc/microMiner & \cite{Muntoni21_microminer} & $\blacksquare$ \\
        MTO & \textit{microTOM} & di-unipi-socc/microTOM & \cite{Soldani23_microtom} & $\square$\\
        PRO & \textit{Prophet} & cloudhubs/prophet & \cite{Bushong21_prophet} & $\blacksquare$ \\
        PR2 & \textit{Prophet2} & cloudhubs/prophet2 & \cite{schiewe2022advancing} & $\square$\\
        PGS & \textit{protoc-gen-scip} & CUHK-SE-Group/protoc-gen-scip & \cite{fang2023rpcover} & $\square$\\
        RAD & \textit{RAD} & cloudhubs/rad & \cite{Das21_rad} & $\blacksquare$ \\
        RAS & \textit{RAD-source} & cloudhubs/rad-source & \cite{Das21_rad} & $\blacksquare$ \\
        \midrule
        \textbf{ID} & \textbf{Name} & \textbf{Website} & & \\
        CMA & \textit{Contextmap} & \url{contextmap.io} & & $\blacksquare$ \\
        
        \bottomrule
    \end{tabular}
\end{table}

The analysis of the selected articles from the formally published literature and resources from the gray literature resulted in the identification of five tools that are in the scope of our study and were not identified by Bakhtin et al. previously~\cite{Bakhtin23_tools_study}.
Together with the tools identified by them that fit the selection criteria of our study, 13 tools are in the scope of our study.
Table~\ref{tbl:tools} presents the final list of twelve open-source and one proprietary tools.

\noindent\fbox{%
    \parbox{0.97\linewidth}{%
    \begin{description}
        \item [\textbf{Answer to RQ1:}] 
        Via the multivocal literature review, we identified 13 static analysis tools to recover the software architecture of microservice applications. Twelve of the tools are open source, one is proprietary. Nine tools are further evaluated in the comparison of tools.
    \end{description}%
    }
}

\vspace{4mm}

The tools' source code and corresponding publications were analyzed in detial to identify characteristics they are expected to detect (step \Circled{2} in Figure~\ref{fig:methodology}).
We also attempted to run each of the identified tools to obtain results of the architecture recovery (step \Circled{4} in Figure~\ref{fig:methodology}).
In the following, each tool is introduced and the process of executing them is described.
Where tools were excluded from the further comparison study, the reasons are given in detail.


\paragraph{\textbf{authz-flow-analysis}}
Abdelfattah et al.~\cite{abdelfattah2023towards} proposed a human-centric but tool-supported method to analyze the access rights of microservice applications.
Specifically, the tool \textit{authz-flow-analysis} detects all endpoints in a microservice application and determines, which of the operations create, read, update, or delete (CRUD) is associated with each endpoint.
According to the authors, the authorization security policies in a microservice system are based on CRUD operations, and their automatic detection therefore eases the analysis process for a human.
The tool analyzes an application's source code and detects endpoints and the CRUD operation that can be performed via them.
From this information, the application's call graph is constructed.

The detection of CRUD operations in \textit{authz-flow-analysis} is built on top of the identification of endpoints via a library called \texttt{source-code-parser} that creates a language-agnostic abstract syntax tree of the analyzed application and that is also used in another tool by the same authors, \textit{Prophet2}.
In the examination of the other tool, we succeeded in running the functionality of this library but not the rest of the analysis.

This tool is the only one in the list of identified tools that has the detection of CRUD operations in its extraction scope.
We could not have compared its performance to others and therefore did not proceed to create the ground truth needed for this characteristic.


\paragraph{\textbf{Attack Graph Generator}}
Ibrahim et al.~\cite{Ibrahim19_attack-graph-generator} presented their tool \textit{Attack Graph Generator}, which automatically generates attack graphs for microservice applications.
Architecture extraction is performed as preliminary for the generation of the attack graphs, i.e., only a means to perform other analysis.
The tool's approach is based on methods from the field of computer networks. 

For this study, only one of the tool's three main components, the ``Topology Parser'', is relevant.
It parses Docker Compose files to extract the analyzed applications' architecture, which is then used in the further analysis that is not relevant for this study.
The Topology Parser extracts components and connections of analyzed applications.
The tool should also detect privileges granted to some Docker containers, however, we did not observe any indications of this in the results.

The tool's repository contains a shell script that installs all required dependencies and then executes a Python script that runs the main functionality.
The documentation states, that the tool was only tested on a virtual machine running Ubuntu 16.04 LTS.
Unfortunately, this operating system has reached its end of service since 2021 and a virtual machine that we set up with it could not execute programs vital for running the \textit{Attack Graph Generator}.
We instead attempted to run the tool on a MacOS based machine, which resulted in errors.
However, a minor debugging effort revealed that the crashes had occurred after the Topology Parser component had finished its execution.
The creation of the topology graph of the analyzed application had finished successfully.
We accepted the failing of the part of the analysis that is not relevant for our study and proceeded with the created topology graphs.
Since the setup described in the tool's documentation could not be replicated, the results might have been affected, but no indications of this were observed and the generated topology graphs show meaningful results.


\paragraph{\textbf{Code2DFD}} presented by Schneider and Scandariato~\cite{Schneider23_code2dfd} is an approach and tool for the extraction of security-rich dataflow diagrams (DFDs) from the source code of microservice applications.
The extracted DFDs depict the architecture consisting of nodes and edges, as well as additional annotations that represent system properties such as implemented security mechanisms, deployment information, and other.
The approach is based on the detection of keywords in the analyzed applications' source code.
Detected keywords serve as evidence for proving the existence of the characteristics they implement.
Based on this technique, the approach can also create traceability information for the model items, i.e., the place of the evidence for the item's existence in the source code.
For the work presented in this paper, the characteristics components, connections, and endpoints are relevant and can be compared to other tools.

\textit{Code2DFD} is implemented in Python and available on GitHub.
After downloading the source code, the tool is executed via the terminal by providing the link to the analyzed application's GitHub repository. 
Alternatively, applications can be analyzed locally if the source code is available there, and the tool can also be run as a Flask server and the architecture extraction triggered via API requests.
\textit{Code2DFD}'s repository contains instructions on the usage of the different possibilities.
We encountered no obstacles in the process of running the tool and it can analyze all applications in the used dataset.


\paragraph{\textbf{ContextMap}}
is the only proprietary tool that we identified.
It is marketed as an ``automated documentation solution'' for software development, intended to solve issues stemming from manual documentation.
The documentation is centered around architecture recovery.
The tool is made to be integrated into CI/CD pipelines and in this case scans the application at every commit.
A dashboard visualizes the recovered architecture and provides information about the development process such as releases, deployments, development team structure, and similar.
The tool only works for applications using Maven as a build system.

Aside from the commercial version, \textit{Contextmap} also offers a free ``Community'' version, which has some limits on the size of analyzable applications, number of users, etc., but offers the full functionality.
It can be obtained and installed as a Docker image from Docker Hub without any problems. 
To analyze an application, the \texttt{pom.xml} file of each individual microservice needs to be modified by adding \textit{ContextMap} as a plugin.
The architecture extraction is initiated via the terminal for every microservice individually (in the case that the tool is not integrated into a CI/CD pipeline) and the results are shown in a dashboard when running the Docker image locally.

We contacted the \textit{ContextMap} team because the tool's documentation states, that the tool should only require changing the root Maven file for multi-module projects and that the analysis does not need to be performed for each microservice individually.
Further, we did not see any connections being detected.
We received an answer giving us the above-described solution, and also the information that the tool only detects connections implemented via Spring's \textit{FeignClient} or \textit{HttpExchange}.
The detection of other connections can be enabled by manually annotating clients in the source code with a custom annotation.
But most connections in the applications in our dataset use \textit{FeignClient} or \textit{HttpExchange}, and we deemed a manual annotation to not be in line with the executed study.


\paragraph{\textbf{MicroDepGraph}} is presented by Rahman et al.~\cite{Rahman19_microdepgraph} in the context of creating a dataset of microservice applications.
It visualizes the dependencies between the microservices of microservice applications to show a graph of components and connections.
The tool analyzes Docker Compose files to recover the required information, and can detect API calls between services directly in Java source code.

\textit{MicroDepGraph} is provided as a Java project on GitHub.
After building it with Maven, it can be executed on applications locally, i.e., the source code of applications to be analyzed need to be cloned from GitHub or obtained otherwise.

When we executed the tool, it crashed when trying to access a Neo4J database, which we had not set up because the documentation does not ask for it.
However, the tool saves the recovered architecture (called ``topology graph'' there) of the analyzed application as an .svg file before attempting to access the database.
According to the tool's corresponding paper, the information contained in the .svg file and database are the same.
Therefore, we did not proceed with trying to solve this issue but see the creation of the .svg file as successful execution of the tool and based our analysis on it.


\paragraph{\textbf{MicroGraal}}

Hutcheson et al. \cite{kozak2024micrograal} proposed a framework to perform static reconstruction of microservice systems based on compiled bytecode using the GraalVM native image.
The use of bytecode instead of source code or deployment information is what makes this approach distinct from most others.
The authors note, that applications' source code might not be available in certain scenarios due to security or proprietary rights reasons.
\textit{MicroGraal} focuses on Java applications using the Spring framework.

Applications to be analyzed need to be compiled with the authors' custom version of the GraalVM native image, which is also available on GitHub.
\textit{MicroGraal} is then executed with the output of the compilation and the list of microservice base package names as input and can detect instances of Java annotations, leading to the recovery of endpoints and REST calls of the microservices.
A ``communication graph'' and a ``system context map'' are created from the detected information.

We were successful in executing the tool on the application \textit{Train Ticket}~\footnote{\url{https://github.com/FudanSELab/train-ticket/}}, thereby replicating the validation reported by the authors in their publication.   
However, the tool produced empty results for all applications in the dataset we used in our study.
All applied debugging efforts could not resolve this issue.


\paragraph{\textbf{microMiner}} presented by Muntoni et al.~\cite{Muntoni21_microminer} is a hybrid approach to recover microservice applications' architecture via Kubernetes deployment files.
The static part of the analysis is the first step in the approach and can be executed independently, therefore, we included it in our study.
The static mining creates a ``partial topology graph'', which contains the application's microservices. 
Connections between them are only extracted in later analysis steps dynamically and are therefore not considered in our comparison.

The tool is implemented in Python.
After cloning the tool's repository, it can be run on local applications' source code directly.
For applications on GitHub, their source code needs to be cloned prior to the analysis.
We faced some difficulties with suppressing the tool's dynamic analysis part, but received support from the authors upon contacting them.
Still, some minor issues occurred in the parsing of Kubernetes files, based on some text comprehension commands that did not consider all legal structures of Kubernetes files.
We adjusted the tool slightly and could then run it successfully.
Also, we leveraged results from previous work~\cite{Cao24_catma} in which some applications in the dataset were extended to be deployed via Kubernetes.
The Kubernetes manifests created for this were added to the dataset to make more applications analyzable with \textit{microMiner}.

The output of \textit{microMiner} is a YAML file following the \textit{microTOSCA} format~\footnote{https://github.com/di-unipi-socc/microTOSCA}, an extension of the TOSCA modelling standard specifically for microservice applications.
The \textit{microTOSCA} files can be visualized with another tool, \textit{microFreshener}~\footnote{https://github.com/di-unipi-socc/microFreshener}. 
This visualization, however, was broken when we ran \textit{microFreshener}.
Instead of investigating this issue, we used the YAML files directly to perform our analysis -- a less convenient process but possible to do manually since the format lists components in a structured way.


\paragraph{\textbf{microTOM}} presented by Soldani et al.~\cite{Soldani23_microtom} is a hybrid tool that recovers microservice applications' architecture from their Kubernetes deployment files.
It was developed by the same authors as \textit{microMiner} (see paragraph above) and is closely related to it, following the same detection technique and adding more elaborate information to the created graph.
The static analysis part of the approach, i.e., the parsing of Kubernetes deployment files, follows the same process as in \textit{microMiner}.
One of the authors confirmed, that the static part of the analysis is equivalent between both tools and that the same results are achieved.
Consequently, we did not consider this tool in the following comparison, since the evaluated part of it would have been a duplicate.
The interested reader may refer to both tools to compare the dynamic parts of the analysis.


\paragraph{\textbf{Prophet}} by Bushong et al.~\cite{Bushong21_prophet} analyzes the applications' source code to recover endpoints and connections between them.
Apart from detecting Java annotations to this end, it also relies on ``enterprise standards'', e.g., by detecting function names that are commonly chosen as identifiers by developers.
The tool also has the applications' components in its extraction scope.
It detects them via the folder structure of the analyzed code (which can be local or on GitHub).
Specifically, all folders in the root of the provided directory are detected as components of the application.
An additional extracted characteristic is a ``context map'', which is created by detecting all classes in the application and identifying those that act as entities, but no further information is provided as to what such a context map looks like or represents.
Overall, the specific workings of the tool are not clear by either the publication or the code repository.

\textit{Prophet} is built on top of \textit{RAD} / \textit{RAD-source} (see below) by the same team that created these tools.
It is available as a library and as a microservice wrapper around that library, which accepts a POST request with a path to the analyzed project as a parameter.
It analyzes the code locally, i.e., if the provided path is a GitHub repository, this will be cloned first.


\paragraph{\textbf{Prophet2}} is a migration of the \textit{Prophet} tool (see above) to the Rust programming language presented by Schiewe et al.~\cite{schiewe2022advancing}.
It further enhances the analysis technique from being Java-specific to language-agnostic by first representing the analyzed application as a language-agnostic abstract syntax tree (LAAST).
In these, different characteristics can theoretically be identified, which is demonstrated in the paper with a case-study on two applications.
The tool implemented for this can detect components, connections, and endpoints, where the detection is largely based on leveraging naming conventions of identifiers in the code.

The tool is centered around a separate submodule called ``source-code-parser'', which can also be run as a stand-alone service, and is responsible for creating the LAASTs.
We could successfully compile this submodule after some difficulties with outdated versions, and could then generate LAASTs by sending requests containing a list of the files to be analyzed to it.
However, \textit{Prophet2} itself did not compile and could not be fixed by us, despite considerable debugging effort.
The tool's repository contains no documentation or even information on how to run it, all above information stems from our analysis of the source code.
We reached out to the authors of the tool but could not receive any support, because the tool has not been used in a long time and the main developers of the tool have moved on from academia.
We decided that abandoning the tool from our study was reasonable, based on the invested effort and inability to receive further support.


\paragraph{\textbf{protoc-gen-scip}} is a plugin for the \texttt{protoc} compiler developed by Fang et al.~\cite{fang2023rpcover}.
It targets microservice applications using the RPC protocol for inter-service communication instead of REST API calls.

The approach uses SCIP~\footnote{https://github.com/sourcegraph/scip} to represent code dependencies of individual microservices via language-specific tools (e.g., \texttt{scip-go}, \texttt{scip-java}) that need to be installed.
These results are used as input for \textit{protoc-gen-scip}, which first creates a single SCIP index for all microservices by matching RPC calls between services and then creates a system overview as a graph out of it.

All applications in the dataset used in this paper use REST calls for communication between services.
The authors of \textit{protoc-gen-scip} themselves note, that there is a lack of available applications using the RPC protocol.
For an evaluation in their paper, they use one open-source application and one application they created for this purpose.
We decided to abandon this tool from the further study, because the creation of a ground truth for a number of new applications would take a large effort and not yield a fair comparison of tools, since there would be no overlap between the applications analyzed be \textit{protoc-gen-scip} and those analyzed by any other tool.


\paragraph{\textbf{RAD}} is a tool proposed by Das et al.~\cite{Das21_rad} that is part of a larger approach to construct a role-based access control model (i.e., a mapping of the access roles required to interact with endpoints)and determine inconsistencies in it.
The architecture recovery is the initial step in the process.
The tool detects endpoints implemented with certain Java annotations and identifies connections between them.
The resulting architecture is then further used for the access control analysis in a human-in-the-loop approach. 

The tool is provided as a Java project on GitHub and can be built with Maven.
Then, it can be run as a microservice and the analysis be executed by sending an API request to it.
It takes the bytecode of the applications as input, therefore, any application to be analyzed needs to be compiled first.

The tool successfully identified endpoints and connections for the application that was used as an evaluation in the corresponding paper.
For the applications in the dataset used in this paper, however, only endpoints were detected.

\paragraph{\textbf{RAD-source}}
The authors of \textit{RAD} (see paragraph above) provide \textit{RAD-source} as a second tool in the same publication~\cite{Das21_rad}.
It is an alternative to \textit{RAD} that takes the analyzed applications' source code as input instead of the compiled bytecode.
The recovered architecture should be the same as that of \textit{RAD}, and the integration of \textit{RAD-source} in the larger approach of access control analysis is the same as well.
The tool is also provided as a Java project that is built and then run as a microservice that takes API requests to trigger the analysis.
In contrast to \textit{RAD}, \textit{RAD-source} identified some connections in the analyzed applications.


\vspace{3mm}
\noindent\fbox{%
    \parbox{0.97\linewidth}{%
    The description of the identified tools and the characteristics they extract serve as the answer to RQ2. In summary:
    \begin{description}
        \item [\textbf{Answer to RQ2:}] There are five characteristics in addition to components and connections extracted by one or more of the 13 identified tools: \textit{endpoints} (extracted by eight tools), \textit{CRUD operations} (one tool), \textit{attack graphs} (one tool), \textit{security and other properties} (one tool), and \textit{context maps} (two tools).
    \end{description}%
    }
}

\subsection{Extracted Characteristics}
\label{sub:extracted_characteristics}

The analysis of the identified tools that the above description of the tools is based on also yielded the information for a tool characterization matrix (step~\Circled{2} in Figure~\ref{fig:methodology}), indicating for each tool which characteristics it is intended to extract from the applications it is used to analyze.
In the remainder of this article, the group of characteristics a tool should extract is called its \textit{extraction scope}.
Table~\ref{tbl:characteristics_matrix} presents the tool characterization matrix, showing each tool's extraction scope.

\begin{table}[!ht]
    \footnotesize
    \centering
	\caption{Tool characterization matrix. }
	\label{tbl:characteristics_matrix}
    \begin{tabular}{p{0.9cm}|P{0.5cm}P{0.5cm}P{0.5cm}P{0.5cm}P{0.5cm}P{0.5cm}P{0.5cm}}
        \toprule
         & \multicolumn{7}{c}{\textbf{Characteristic}} \\
        \cline{2-8}
         \textbf{Tool} & \rotatebox[origin=l]{90}{Components} & \rotatebox[origin=l]{90}{Connections} & \rotatebox[origin=l]{90}{\parbox{2.2cm}{Endpoints}} & \rotatebox[origin=l]{90}{\parbox{2.2cm}{CRUD \\operations}} & \rotatebox[origin=l]{90}{Attack graphs} & \rotatebox[origin=l]{90}{\parbox{2.2cm}{Security and \\other properties}} & \rotatebox[origin=l]{90}{\parbox{2.2cm}{Context map}} \\
        \midrule
        AFA & $\square$ & $\blacksquare$ & $\blacksquare$ & $\blacksquare$ & $\square$ & $\square$ & $\square$ \\
        AGG & $\blacksquare$ & $\blacksquare$ & $\square$ & $\square$ & $\blacksquare$ & $\square$ & $\square$ \\
        C2D & $\blacksquare$ & $\blacksquare$ & $\blacksquare$ & $\square$ & $\square$ & $\blacksquare$ & $\square$ \\
        MDG &  $\blacksquare$ & $\blacksquare$ & $\square$ & $\square$ & $\square$ & $\square$ & $\square$ \\
        MGR & $\square$ & $\blacksquare$ & $\blacksquare$ & $\square$ & $\square$ & $\square$ & $\blacksquare$ \\
        MMI & $\blacksquare$ & $\square$ & $\square$ & $\square$ & $\square$ & $\square$ & $\square$ \\
        MTO & $\blacksquare$ & $\square$ & $\square$ & $\square$ & $\square$ & $\square$ & $\square$ \\
        PRO & $\blacksquare$ & $\blacksquare$ & $\blacksquare$ & $\square$ & $\square$ & $\square$ & $\blacksquare$ \\
        PR2 & $\blacksquare$ & $\blacksquare$ & $\blacksquare$ & $\square$ & $\square$ & $\square$ & $\square$ \\
        PGS & $\square$ & $\blacksquare$ & $\square$ & $\square$ & $\square$ & $\square$ & $\square$ \\
        RAD & $\square$ & $\blacksquare$ & $\blacksquare$ & $\square$ & $\square$ & $\square$ & $\square$ \\
        RAS & $\square$ & $\blacksquare$ & $\blacksquare$ & $\square$ & $\square$ & $\square$ & $\square$ \\
        CMA & $\square$ & $\blacksquare$ & $\blacksquare$ & $\square$ & $\square$ & $\square$ & $\square$ \\

        \bottomrule
    \end{tabular}
\end{table}

Table~\ref{tbl:characteristics_matrix} shows, that the most common characteristic in the tools' extraction scopes are connections between components. 
Eleven tools are intended to extract it, only two are not.
Interestingly, only seven of the tools are meant to extract those components in the first place.
The others require the list of components to be provided as input or the tool to be executed for each component individually, which we do not consider to be an extraction of the components.
There are five characteristics in the tools' extraction scopes beyond the basic architecture (which are the components and connections).
Seven tools extract the microservices' REST endpoints, which give further details about the connections made via them.
Two closely related tools detect access control inconsistencies. Two other related tools provide a Context map of all system entities. 
The other four characteristics (CRUD operations, attack graphs, security and other system properties, and abstract syntax tree) are in the extraction scope of a single tool each.

\noindent\fbox{%
    \parbox{0.97\linewidth}{%
    Table~\ref{tbl:characteristics_matrix} provides the answer to RQ3:
    \begin{description}
        \item [\textbf{Answer to RQ3:}] The most commonly considered characteristic by the tools are connections between components (eleven out of 13 tools extract them).
        The systems' components (seven tools) and REST endpoints (seven tools) are also prevalent in the extraction scopes of the tools.
        Other characteristics are considered by not more than two tools for the same characteristic.
    \end{description}%
    }
}

\section{Comparison of Tools}
\label{sec:comparison}

Comparing the tools' accuracy in performing their specific architecture extraction is a novelty of this work that fills a gap in the related literature published so far.
As discussed in Section~\ref{sub:identified_tools}, the two tools \textit{authz-flow-analysis} and \textit{microTOM} will not be considered in this comparison, since they realize the architectural reconstruction with other identified tools and do not qualify as separate tools for the sake of this study.
The tools \textit{Prophet2} and \textit{protoc-gen-scip} are also excluded, as described.

After executing the nine tools that are considered in the comparison on all applications in the dataset, we manually quantified the results following the methodology presented in Section~\ref{sec:methodology}.
The following sections present the results of this process per extracted characteristic and compare those tools that have that characteristic in their extraction scope.
We note that judging the results depends highly on the usage scenario and that the reader should rely on the raw metrics rather than our description of what is considered a high or low result when considering adopting one of the tools for a specific scenario.
Nevertheless, we use such classifying terms to better show the comparison of the tools' observed correctness without further context.

\begin{table}
    \footnotesize
    \centering
    \caption{Tools' extraction results for the characteristic (a)\textit{components}, (b) \textit{connections}, and (c) \textit{endpoints}.\\\textbf{GT} = number of individual characteristics in the applications that the tool should be able to analyze; \textbf{TP}/\textbf{FP}/\textbf{FN} = observed true positives / false positives / false negatives; \textbf{P}/\textbf{R}/\textbf{F1} = calculated precision / recall / F1-score; Dashes (---) = characteristic is not in tool's extraction scope.}
    \begin{subtable}{\textwidth}
        \centering
        \caption{Components}
        \label{tbl:results_components}
        \begin{tabular}{l|r|rrr|rrr}
            \toprule
            \textbf{Tool} & \textbf{GT} & \textbf{TP} & \textbf{FP} & \textbf{FN} & \textbf{P} & \textbf{R} & \textbf{F1}\\
            \midrule 
            AGG & 144  & 123 & 41 & 21 & 0.75 & 0.85 & 0.8 \\
            C2D & 182 & 178 & 5 & 4 & 0.97 & \textbf{0.98} & \textbf{0.98} \\
            MDG & 110 & 84 & 0 & 26 & \textbf{1.0} & 0.76 & 0.87 \\
            MGR & --- & --- & --- & --- & --- & --- & --- \\
            MMI & 57 & 32 & 1 & 25 & 0.97 & 0.56 & 0.71 \\
            PRO & 182 & 27 & 21 & 155 & 0.56 & 0.15 & 0.23 \\
            RAD & --- & --- & --- & --- & --- & --- & --- \\
            RAS & --- & --- & --- & --- & --- & --- & --- \\
            CMA & --- & --- & --- & --- & --- & --- & --- \\
            \bottomrule
        \end{tabular}
    \end{subtable}
    \hfill
    \vspace{1mm}
    \begin{subtable}{\textwidth}
        \centering
        \caption{Connections}
        \label{tbl:results_connections}
        \begin{tabular}{l|r|rrr|rrr}
            \toprule
            \textbf{Tool} & \textbf{GT} & \textbf{TP} & \textbf{FP} & \textbf{FN} & \textbf{P} & \textbf{R} & \textbf{F1}\\
            \midrule 
            AGG & 324 & 279 & 432 & 45 & 0.39 & \textbf{0.86} & 0.54 \\
            C2D & 385 & 320 & 27 & 65 & 0.92 & 0.83 & \textbf{0.87} \\
            MDG & 245 & 124 & 2 & 121 & \textbf{0.98} & 0.51 & 0.67 \\
            MGR & 173 & 0 & 0 & 173 & n/a & 0 & 0 \\
            MMI & --- & --- & --- & --- & --- & --- & --- \\
            PRO & 385 & 4 & 2 & 381 & 0.67 & 0.01 & 0.02 \\
            RAD & 259 & 0 & 0 & 259 & n/a & 0 & 0 \\
            RAS & 385 & 7 & 2 & 378 & 0.78 & 0.02 & 0.04 \\
            CMA & 206 & 0 & 0 & 206 & n/a & 0 & 0 \\
            \bottomrule
        \end{tabular}
    \end{subtable}
    \hfill
    \vspace{1mm}
    \begin{subtable}{\textwidth}
        \centering
        \caption{Endpoints}
        \label{tbl:results_endpoints}
        \begin{tabular}{l|r|rrr|rrr}
            \toprule
            \textbf{Tool} & \textbf{GT} & \textbf{TP} & \textbf{FP} & \textbf{FN} & \textbf{P} & \textbf{R} & \textbf{F1}\\
            \midrule 
            AGG & --- & --- & --- & --- & --- & --- & --- \\
            C2D & 160 & 86 & 13 & 74 & 0.87 & 0.54 & 0.66 \\
            MDG & --- & --- & --- & --- & --- & --- & --- \\
            MGR & 121 & 0 & 0 & 121 & n/a & 0 & 0 \\
            MMI & --- & --- & --- & --- & --- & --- & --- \\
            PRO & 160 & 0 & 0 & 160 & n/a & 0 & 0 \\
            RAD & 132 & 90 & 6 & 42 & \textbf{0.94} & \textbf{0.68} & \textbf{0.79} \\
            RAS & 160 & 86 & 11 & 74 & 0.89 & 0.54 & 0.67 \\
            CMA & 91 & 0 & 0 & 91 & n/a & 0 & 0 \\
            \bottomrule
        \end{tabular}
    \end{subtable}
    \vspace{1mm}
    \begin{subtable}{\textwidth}
        \centering
        \caption{All characteristics in tool's extraction scope; \\Filled box under \textbf{Cp}/\textbf{Cn}/\textbf{E} = components / connections / endpoints are in tool's extraction scope.}
        \label{tbl:results_all_characteristics}
        \begin{tabular}{l|P{0.4cm}P{0.4cm}P{0.4cm}|r|rrr|rrr}
            \toprule
        \textbf{Tool} & \textbf{Cp} & \textbf{Cn} & \textbf{E} & \textbf{GT} & \textbf{TP} & \textbf{FP} & \textbf{FN} & \textbf{P} & \textbf{R} & \textbf{F1}\\
        \midrule 
        AGG & $\blacksquare$ & $\blacksquare$ & $\square$ & 468 & 402 & 473 &  66 & 0.46 & \textbf{0.86} & 0.60 \\
        C2D & $\blacksquare$ & $\blacksquare$ & $\blacksquare$ & 727 & 584 &  45 & 143 & 0.93 & 0.80 & \textbf{0.86} \\
        MDG & $\blacksquare$ & $\blacksquare$ & $\square$ & 355 & 208 &   2 & 147 & \textbf{0.99} & 0.59 & 0.74 \\
        MGR & $\square$ & $\blacksquare$ & $\blacksquare$ & 295 &   0 &   0 & 295 & n/a  & 0    & 0    \\
        MMI & $\blacksquare$ & $\square$ & $\square$ &  57 &  32 &   1 &  25 & 0.97 & 0.56 & 0.71 \\
        PRO & $\blacksquare$ & $\blacksquare$ & $\blacksquare$ & 727 &  31 &  23 & 696 & 0.57 & 0.04 & 0.08 \\
        RAD & $\square$ & $\blacksquare$ & $\blacksquare$ & 391 &  90 &   6 & 301 & 0.94 & 0.23 & 0.37 \\
        RAS & $\square$ & $\blacksquare$ & $\blacksquare$ & 545 &  93 &  13 & 452 & 0.88 & 0.17 & 0.29 \\
        CMA & $\square$ & $\blacksquare$ & $\blacksquare$ & 297 &   0 &   0 & 297 & n/a  & 0    & 0    \\
        \bottomrule
        \end{tabular}
    \end{subtable}
\end{table}

\subsection{Individual Tools}

Herein, we present the comparative results of the identified tools' extraction accuracy for different characteristics.

\paragraph{\textbf{Components}}
Five of the identified tools extract the analyzed systems' components, i.e., their individual microservices.
Table~\ref{tbl:results_components} shows the observed results.
Three tools (\textit{MicroDepGraph}, \textit{Code2DFD}, and \textit{MicroMiner}) show a high precision with results of 1.0, 0.97, and 0.97 in this metric, respectively.
\textit{Attack Graph Generator} and \textit{Prophet} show higher amounts of false positives, resulting in a precision of 0.75 and 0.56, respectively. 
\textit{Code2DFD} is able to extract almost all components, with only five false negatives out of 182 components in the ground truth, resulting in a recall of 0.98. 
\textit{Attack Graph Generator} and \textit{MicroDepgraph} with a recall of 0.85 and 0.76, respectively, also perform well.
\textit{MicroMiner} shows a recall of 0.56 and \textit{Prophet} only identifies few components (recall of 0.15).
Looking at the F1-score, \textit{Code2DFD} achieves the best result with a value of 0.98.
The next-best tools are \textit{MicroDepGraph} with a score of 0.87 and \textit{Attack Graph Generator} with a score of 0.80, followed by \textit{microMiner} with 0.71.
\textit{Prophet} achieves a low score of 0.23.

In summary, we observed the highest recall and one of the highest precision values for \textit{Code2DFD} for this characteristic, resulting in the best F1-score of 0.98.
\textit{MicroDepGraph} showed no false positives, which resulted in a perfect precision, but a lower recall.
\textit{Attack Graph Generator} shows good results in all metrics, while \textit{microMiner} also has a high precision but low recall.
\textit{Prophet} shows the worst accuracy out of the tools concerning this characteristic.

\noindent\fbox{%
    \parbox{0.97\linewidth}{%
    \begin{description}
        \item [\textbf{Answer to RQ4.1:}] Table~\ref{tbl:results_components} presents the tools' observed accuracy in extracting the analyzed applications' components.
        \textit{Code2DFD} showed the highest F1-score of 0.98, \textit{MicroDepGraph} a score of 0.87, \textit{Attack Graph Generator} a score of 0.80, and \textit{microMiner} a score of 0.71.
    \end{description}%
    }
}

\paragraph{\textbf{Connections}}
Eight of the identified tools have the analyzed applications' connections in their extraction scopes.
Table~\ref{tbl:results_connections} presents the observed results.
It shows that only three tools were able to do so somewhat effectively. 
Three tools (\textit{MicroGraal}, \textit{RAD}, and \textit{ContextMap}) did not detect any connections, resulting in a recall of 0 (and the calculation of the precision is not applicable).
Two other tools only extracted a miniscule amount of connections (\textit{Prophet} showed four true positives resulting in an F1-score of 0.02; \textit{RAD-source} detected 7 connections, resulting in an F1-score of 0.04) and can hence also be considered as not effective in extracting this characteristic.
These results occurred despite us having confirmed that the tools executed correctly.

From the three tools showing better extraction correctness, \textit{MicroDepGraph} and \textit{Code2DFD} showed a high precision (0.98 and 0.92, respectively), and \textit{Code2DFD} and \textit{Attack Graph Generator} a good recall (0.86 and 0.83, respectively). 
With a value of 0.39, the precision of \textit{Attack Graph Generator} is rather low, and with a value of 0.51, the recall of \textit{MicroDepGraph} as well.
These yield a F1-score of 0.87 for \textit{Code2DFD}, 0.67 for \textit{MicroDepGraph}, and 0.54 for \textit{Attack Graph Generator}.

Overall, \textit{MicroDepGraph} shows a high precision but low recall and \textit{Attack Graph Generator} the opposite. 
\textit{Code2DFD} is more balanced and achieves high scores in both metrics.
This results in \textit{Code2DFD} having the highest F1-score.

\noindent\fbox{%
    \parbox{0.97\linewidth}{%
    \begin{description}
        \item [\textbf{Answer to RQ4.2:}] Table~\ref{tbl:results_connections} presents the tools' observed accuracy in extracting the analyzed applications' connections.
        The highest F1-score of 0.87 was observed for \textit{Code2DFD}, followed by \textit{MicroDepGraph} (F1 = 0.67) and \textit{Attack Graph Generator} (F1 = 0.54).
    \end{description}%
    }
}

\paragraph{\textbf{Endpoints}}
Six of the identified tools extract the applications' REST endpoints during analysis, i.e., endpoints over which RESTful API requests can be made to services.
Table~\ref{tbl:results_endpoints} shows the results of the six tools for this characteristic.
Three tools (\textit{MicroGraal}, \textit{Prophet}, and \textit{ContextMap}) failed in detecting any endpoints completely, resulting in a recall of zero and non-applicable precision.
From the other three, we observed a very high precision for \textit{RAD} (0.94) with only six false positives.
It also had a good recall with 0.68.
\textit{RAD-source} had the second-highest precision with a value of 0.89, but a low recall of 0.54.
Similarly, \textit{Code2DFD} showed a high precision (0.87) but low recall, achieving 0.54 in this metric.
Combined into the F1-score, \textit{RAD} scores 0.79, \textit{RAD-source} 0.67, and \textit{Code2DFD} 0.66.

\noindent\fbox{%
    \parbox{0.97\linewidth}{%
    \begin{description}
        \item [\textbf{Answer to RQ4.3:}] Table~\ref{tbl:results_endpoints} presents the tools' observed accuracy in extracting the analyzed applications' endpoints.
        \textit{RAD} achieved the highest F1-score of 0.79.
        \textit{RAD-source} and \textit{Code2DFD} performed comparably well with F1-scores of 0.67 and 0.66, respectively. 
    \end{description}%
    }
}

\paragraph{\textbf{All Characteristics in Extraction Scope}}

Combining the results per characteristic presented above, Table~\ref{tbl:results_all_characteristics} presents again each tool's extraction scope and the observed results for all characteristics in its extraction scope.
In other words, the table shows each tools' correctness for performing the architecture extraction it is intended to do.
Two tools (\textit{MicroGraal} and \textit{ContextMap}) failed completely, and three more tools (\textit{Prophet}, \textit{RAD}, and \textit{RAD-source}) also achieved low scores overall when looking at the F1-score as a combination of precision and recall (scores of 0.10, 0.38, and 0.29, respectively).
The other three tools exhibited similar results as for the components and connections: \textit{Attack Graph Generator} showed a high number of true positives but also false positives and a resulting highest recall of 0.86 but low precision; \textit{MicroDepGraph} showed a low number of false positives but also less true positives resulting in the highest precision of 0.99 but low recall; and \textit{Code2DFD} showed more balance between the measures and a resulting good precision, good recall, and overall highest F1-score with some distance of 0.86.

\paragraph{\textbf{Execution Time}}

\begin{table}
    \footnotesize
    \centering
	\caption{Tools' execution times for analyzing the indicated number of applications individually. Average over ten execution runs. Tool with dashes (---) requires manual steps to a large degree.}
	\label{tbl:execution_times}
    \begin{tabular}{l|c|S[table-format=2.2]S[table-format=2.2]S[table-format=2.2]}
        \toprule
         & \textbf{Analyzed} & \multicolumn{3}{c}{\textbf{Time per App. [sec]}} \\
        \cline{3-5}
        \textbf{Tool} & \textbf{Apps.} & \textbf{Min.} & \textbf{Max.} & \textbf{Avg.} \\
        \midrule 
        AGG &  9 & 1.2 & 1.5 & 1.3 \\
        C2D & 17 & 6.7 & 7.8 & 7.3 \\
        MDG & 10 & 1.3 & 1.6 & 1.5 \\
        MGR & 10 & 50 & 144 & 60 \\
        MMI &  5 & 0.20 & 0.22 & 0.20 \\
        PRO & 17 & 0.16 & 0.17 & 0.17 \\
        RAD & 13 & 0.03 & 0.03 & 0.03 \\
        RAS & 17 & 0.04 & 0.04 & 0.04 \\
        CMA & 10 & {---} & {---} & {---} \\
        \bottomrule
    \end{tabular}
\end{table}

The tools' execution times for performing the architecture extraction were measured to allow an assessment of their suitability to be integrated in time-sensitive analysis settings such as fast-paced CI/CD pipelines.
Table~\ref{tbl:execution_times} presents the results.

\textit{RAD} showed the quickest execution times with an average of 0.03 seconds per application over ten execution runs, directly followed by \textit{RAD-source} with 0.04.
\textit{Prophet} and \textit{microMiner} also average well below a second per application, with 0.17 and 0.20, respectively.
\textit{Attack Graph Generator} (1.3) and \textit{MicroDepgraph} (1.5) average slightly more than one second, while code2DFD takes 7.3 seconds per application on average.
With an average of one minute per application, \textit{microGraal} showed the longest execution times in our study.

\subsection{Combinations of Tools}

To investigate whether multiple tools could be combined to exceed the extraction correctness of the individual tools, the results of different combinations of tools were combined and the scores they achieved were assessed.
Based on the results of the individual tools, only a small subset of all possible permutations of tools are reasonable to investigate.

For the characteristics components and connections, the tools \textit{Attack Graph Generator}, \textit{Code2DFD}, \textit{microMiner}, and \textit{MicroDepGraph} are the only ones that achieved good results.
A comparison of the detailed results shows, that \textit{microMiner} does not provide any additional value over the others in terms of detecting characteristics that the others did not and is therefore omitted, leaving \textit{Attack Graph Generator}, \textit{Code2DFD}, and \textit{MicroDepGraph} for the tool combinations.
For the detection of endpoints, the results of \textit{Code2DFD}, \textit{RAD}, and \textit{RAD-source} are all high and distinct, i.e., they were successful in the detection of different endpoints.
The permutations of all three are hence assessed for these characteristics.
Finally, for the extraction of all characteristics, we combine the best tool combinations for the detection of each characteristic per metric and compare them against a combination of all seven tools that were part of the study and produced results for any characteristic.
Table~\ref{tbl:tool_combinations} presents the combinations of tools considered in the evaluation, resulting from the above considerations.

\begin{table}
    \footnotesize
    \centering
	\caption{Combinations of tools that were evaluated for each characteristic. \\\textbf{Cp} = components; \textbf{Cn} = connections; \textbf{E} = endpoints.}
	\label{tbl:tool_combinations}
    \begin{tabular}{llccccccc}
        \toprule
         & & \multicolumn{7}{c}{\textbf{Included Tools}} \\
         \cline{3-9}
        \textbf{Characteristics} & \textbf{ID} & AGG & C2D & MDG & MMI & PRO & RAD & RAS \\
        \midrule
        \textbf{Cp}, \textbf{Cn}, \textbf{E} & All & $\blacksquare$ & $\blacksquare$ & $\blacksquare$ & $\blacksquare$ & $\blacksquare$ & $\blacksquare$ & $\blacksquare$ \\
         & Best\textsubscript{P} & \multicolumn{7}{c}{\textit{Combination with highest P for each characteristic}} \\
         & Best\textsubscript{R} & \multicolumn{7}{c}{\textit{Combination with highest R for each characteristic}} \\
         & Best\textsubscript{F1} & \multicolumn{7}{c}{\textit{Combination with highest F1 for each characteristic}} \\
        \midrule
        \textbf{Cp} & ACM & $\blacksquare$ & $\blacksquare$ & $\blacksquare$ & $\square$ & $\square$ & $\square$ & $\square$ \\
         & AC & $\blacksquare$ & $\blacksquare$ & $\square$ & $\square$ & $\square$ & $\square$ & $\square$ \\
         & AM & $\blacksquare$ & $\square$ & $\blacksquare$ & $\square$ & $\square$ & $\square$ & $\square$  \\
         & CM  & $\square$ & $\blacksquare$ & $\blacksquare$ & $\square$ & $\square$ & $\square$ & $\square$ \\
        \midrule
        \textbf{Cn} & ACM & $\blacksquare$ & $\blacksquare$ & $\blacksquare$ & $\square$ & $\square$ & $\square$ & $\square$ \\
         & AC & $\blacksquare$ & $\blacksquare$ & $\square$ & $\square$ & $\square$ & $\square$ & $\square$ \\
         & AM & $\blacksquare$ & $\square$ & $\blacksquare$ & $\square$ & $\square$ & $\square$ & $\square$  \\
         & CM  & $\square$ & $\blacksquare$ & $\blacksquare$ & $\square$ & $\square$ & $\square$ & $\square$ \\
        \midrule 
        \textbf{E} & CRR  & $\square$ & $\blacksquare$ & $\square$ & $\square$ & $\square$ & $\blacksquare$ & $\blacksquare$ \\
         & CRD  & $\square$ & $\blacksquare$ & $\square$ & $\square$ & $\square$ & $\blacksquare$ & $\square$ \\
         & CRS  & $\square$ & $\blacksquare$ & $\square$ & $\square$ & $\square$ & $\square$ & $\blacksquare$ \\
         & RDS  & $\square$ & $\square$ & $\square$ & $\square$ & $\square$ & $\blacksquare$ & $\blacksquare$ \\
        \bottomrule
    \end{tabular}
\end{table}

\begin{table}
    \footnotesize
    \centering
    \caption{Tool combinations' extraction results for the characteristic (a)\textit{components}, (b) \textit{connections}, (c) \textit{endpoints}, and (d) \textit{all characteristics}. \\\textbf{GT} = number of individual characteristics in the applications that the combination should be able to analyze; \textbf{TP}/\textbf{FP}/\textbf{FN} = observed true positives / false positives / false negatives; \textbf{P}/\textbf{R}/\textbf{F1} = calculated precision / recall / F1-score.}
    \label{tbl:combination_results_all_subtables}
    \begin{subtable}{\textwidth}
        \centering
        \caption{Components}
        \label{tbl:combination_results_components}
        \begin{tabular}{l|r|rrr|rrr}
            \toprule
            \textbf{Tool Combination} & \textbf{GT} & \textbf{TP} & \textbf{FP} & \textbf{FN} & \textbf{P} & \textbf{R} & \textbf{F1}\\
            \midrule 
            ACM\textsubscript{OR}  & 182 & 179 & 46 &  3 & 0.80 & \textbf{0.98} & 0.88 \\
            ACM\textsubscript{AND} & 182 & 145 &  0 & 37 & \textbf{1.00} & 0.80 & 0.89 \\
            AC\textsubscript{OR}   & 182 & 179 & 46 &  3 & 0.80 & \textbf{0.98} & 0.88 \\
            AC\textsubscript{AND}  & 182 & 158 &  0 & 24 & \textbf{1.00} & 0.87 & 0.93 \\
            AM\textsubscript{OR}   & 144 & 125 & 41 & 19 & 0.75 & 0.87 & 0.81 \\
            AM\textsubscript{AND}  & 144 &  82 &  0 & 62 & \textbf{1.00} & 0.57 & 0.73 \\
            CM\textsubscript{OR}   & 182 & 178 &  5 &  4 & 0.97 & 0.98 & \textbf{0.98} \\
            CM\textsubscript{AND}  & 182 & 153 &  0 & 29 & \textbf{1.00} & 0.84 & 0.91 \\
            \bottomrule
        \end{tabular}
    \end{subtable}
    \hfill
    \vspace{1mm}
    \begin{subtable}{\textwidth}
        \centering
        \caption{Connections}
        \label{tbl:combination_results_connections}
        \begin{tabular}{l|r|rrr|rrr}
            \toprule
            \textbf{Tool Combination} & \textbf{GT} & \textbf{TP} & \textbf{FP} & \textbf{FN} & \textbf{P} & \textbf{R} & \textbf{F1}\\
            \midrule 
            ACM\textsubscript{OR}  & 385 & 373 & 412 &  12 & 0.48 & \textbf{0.97} & 0.64 \\
            ACM\textsubscript{AND} & 385 & 208 &   0 & 177 & \textbf{1.00} & 0.54 & 0.70 \\
            AC\textsubscript{OR}   & 385 & 373 & 412 &  12 & 0.48 & \textbf{0.97} & 0.64 \\
            AC\textsubscript{AND}  & 385 & 279 &   0 & 106 & \textbf{1.00} & 0.72 & 0.84 \\
            AM\textsubscript{OR}   & 325 & 282 &  396 & 43 & 0.42 & 0.87 & 0.56 \\
            AM\textsubscript{AND}  & 325 & 121 &   0 & 204 & \textbf{1.00} & 0.37 & 0.54 \\
            CM\textsubscript{OR}   & 385 & 331 &  29 &  54 & 0.92 & 0.86 & \textbf{0.89} \\
            CM\textsubscript{AND}  & 385 & 219 &   0 & 166 & \textbf{1.00} & 0.57 & 0.73 \\
            \bottomrule
        \end{tabular}
    \end{subtable}
    \hfill
    \vspace{1mm}
    \begin{subtable}{\textwidth}
        \centering
        \caption{Endpoints}
        \label{tbl:combination_results_endpoints}
        \begin{tabular}{l|r|rrr|rrr}
            \toprule
            \textbf{Tool Combination} & \textbf{GT} & \textbf{TP} & \textbf{FP} & \textbf{FN} & \textbf{P} & \textbf{R} & \textbf{F1}\\
            \midrule 
            CRR\textsubscript{OR}  & 160 & 141 & 23 &  19 & 0.86 & \textbf{0.88} & 0.87 \\
            CRR\textsubscript{AND} & 160 &  28 &  2 & 132 & 0.93 & 0.18 & 0.29 \\
            CRD\textsubscript{OR}  & 160 & 132 & 17 &  28 & 0.89 & 0.83 & \textbf{0.85} \\
            CRD\textsubscript{AND} & 160 &  44 &  2 & 116 & 0.96 & 0.28 & 0.43 \\
            CRS\textsubscript{OR}  & 160 & 120 & 20 &  40 & 0.86 & 0.75 & 0.80 \\
            CRS\textsubscript{AND} & 160 &  52 &  2 & 108 & \textbf{0.96} & 0.33 & 0.49 \\
            RDS\textsubscript{OR}  & 160 & 123 & 12 &  37 & 0.91 & 0.77 & 0.83 \\
            RDS\textsubscript{AND} & 160 &  53 &  3 & 107 & 0.95 & 0.33 & 0.49 \\
            \bottomrule
        \end{tabular}
    \end{subtable}
    \vspace{1mm}
    \begin{subtable}{\textwidth}
        \centering
        \caption{All characteristics}
        \label{tbl:combination_results_full}
        \begin{tabular}{l|r|rrr|rrr}
            \toprule
            \textbf{Tool Combination} & \textbf{GT} & \textbf{TP} & \textbf{FP} & \textbf{FN} & \textbf{P} & \textbf{R} & \textbf{F1}\\
            \midrule 
            All\textsubscript{OR}  & 727 & 694 & 509 &  33 & 0.58 & \textbf{0.95} & 0.72 \\
            All\textsubscript{AND} & 727 &  52 &   2 & 675 & 0.96 & 0.07 & 0.13 \\
            Best\textsubscript{P}  & 727 & 489 & 2 &  238 & \textbf{1.00} & 0.76 & 0.80 \\
            Best\textsubscript{R}  & 727 & 694 & 484 &  33 & 0.59 & \textbf{0.95} & 0.73 \\
            Best\textsubscript{F1} & 727 & 650 &  57 &  77 & 0.92 & 0.89 & \textbf{0.91} \\
            \bottomrule
        \end{tabular}
    \end{subtable}
\end{table}

\paragraph{\textbf{Components}}
Table~\ref{tbl:combination_results_components} shows the results for the tool combinations' extraction correctness for the characteristic components.
For each metric, there is a combination that achieves a high score. 
All evaluated ``AND''-combinations exhibit a perfect precision of 1.0, whereas combining tools with an ``OR'' operation yields a high recall of 0.98 for three combinations and 0.87 for AM\textsubscript{OR}.
High F1-scores are also observed, with the combination CM\textsubscript{OR} yielding the best result of 0.98 in F1-score.
Interestingly, the best combination of tools in terms of recall outperforms the best individual tool, \textit{Code2DFD}, by only a single true positive.
\textit{Code2DFD} alone matches the F1-score of the best tool combination in this metric.

\paragraph{\textbf{Connections}}
Table~\ref{tbl:combination_results_connections} shows the results for the combinations of tools for the characteristic connections.
The tool combinations ACM\textsubscript{OR} and AC\textsubscript{OR} exhibit the same results, showing that \textit{MicroDepGraph} does not contribute any additional true positives or false positives over the other two tools.
As for the extraction of components, the tool combinations with ``AND'' show a perfect precision and the combinations with ``OR'' a high recall.
Specifically, AC\textsubscript{OR} achieves a recall of 0.97 as the best combination in this metric.
With this result, the combination of \textit{Attack Graph Generator} and \textit{Code2DFD} shows a substantial improvement over the recall of the best individual tool alone, \textit{Attack Graph Generator}, with a recall of 0.86.
In terms of F1-score, the results are generally lower, mainly caused by many false positives for combinations that contain \textit{Attack Graph Generator} and many false negatives for ``AND''-combinations.
Consequently, the combination CM\textsubscript{OR} shows the highest F1-score of 0.89.
It slightly outperforms the best individual tool, \textit{MicroDepGraph}, which scored 0.88 in this metric.

\paragraph{\textbf{Endpoints}}
Table~\ref{tbl:combination_results_endpoints} presents the results for the tool combinations' extraction correctness for the characteristic endpoints.
All ``AND''-combinations show slightly higher precision than the ``OR''-combinations again, with the highest being 0.96 for CRS\textsubscript{AND}.
The combination of all three considered tools CRR\textsubscript{OR} yields the highest recall of 0.88, a substantial improvement over the best individual tool \textit{RAD} (recall of 0.68).
The F1-score can be marginally improved to 0.85 over the score of 0.79 for \textit{RAD} alone when combining it with \textit{Code2DFD} into CRD\textsubscript{OR}.

\paragraph{\textbf{All characteristics}}
Table~\ref{tbl:combination_results_full} shows the results for the combinations of tools for the full architecture extraction, i.e., components, connections, and endpoints.
The combinations Best\textsubscript{P}, Best\textsubscript{R}, and Best\textsubscript{F1} consist of the best-performing tool or combination of tools for each characteristic in terms of the metric indicated by the combination's ID.
Table~\ref{tbl:optimal_tool_combinations} shows the tool combinations resulting from this selection, which showed to yield the best possible results for each evaluated metric: 

\begin{table}[!ht]
    \centering
    \caption{Optimal tool combinations for each evaluated metric.}
    \label{tbl:optimal_tool_combinations}
    \begin{tabular}{c|ccc}
      \toprule
      \textbf{Combination} & \textbf{Components} & \textbf{Connections} & \textbf{Endpoints} \\
      \midrule
      Best$_\text{P}$ & AC$_\text{AND}$ & AC$_\text{AND}$ & CRS$_\text{AND}$ \\
      Best$_\text{R}$ & AC$_\text{OR}$ & All$_\text{OR}$ & CRR$_\text{OR}$ \\
      Best$_\text{F1}$ & C2D & CM$_\text{OR}$ & CRR$_\text{OR}$ \\
      \bottomrule
    \end{tabular}
    
\end{table}

All tool combinations in Table~\ref{tbl:combination_results_full} consider the full dataset of 727 individual characteristics in the ground truth, i.e., the metrics were not inflated by simply excluding tools that are able to analyze more applications but also produce more FPs.
An almost perfect precision of 1.00 after rounding can be achieved with the tool combination Best\textsubscript{P} (consisting of \textit{Attack Graph Generator}, \textit{Code2DFD}, and \textit{RAD-source}) while still having a reasonably high F1-score of 0.80.
Combining all seven tools that were part of this evaluation with an ``AND''-operator also achieves a high precision of 0.96, but a low recall and F1-score.
The highest recall of a combination of a subset of tools can not outperform the combination of all tools per definition, but also, no other combination of only a subset of tools achieves the same recall combined with a substantially higher precision.
Only a slight improvement can be achieved in this regard.
The set of tools in Best\textsubscript{R} thus contains all tools also contained in All\textsubscript{OR}.
Finally, the best possible overall extraction correctness measured in F1-score is 0.91, showing both a high precision and recall. 
It is a combination of four tools, \textit{Code2DFD}, \textit{MicroDepGraph}, \textit{RAD}, and \textit{RAD-source}.
This combination is an improvement of $\sim$5\% over the best individual tool, \textit{Code2DFD}, which showed an F1-score of 0.86.

\noindent\fbox{%
    \parbox{0.97\linewidth}{%
    \begin{description}
        \item [\textbf{Answer to RQ5:}] 
        Combining multiple tools improves the extraction accuracy in almost all investigated cases.
        For each evaluated performance metric and each characteristic (except of the F1-score for components), a tool combination can be found that outperforms the results of the individual tools.
        A combination of \textit{Code2DFD}, \textit{MicroDepGraph}, \textit{RAD}, and \textit{RAD-source} achieves the highest F1-score of 0.91.
    \end{description}%
    }
}
\section{Discussion}
\label{sec:discussion}

The results presented in Section~\ref{sec:comparison} show the identified tools' accuracy in architecture extraction.
We discuss the results in the following.

\subsection{Tools' Extraction Accuracy}

As a first, general observation concerning the results of the presented comparison of tools, the extent to which valid results could be obtained is limited.
Out of the nine tools considered during the comparison, only six exhibited an extraction correctness that even warrants an inclusion in the comparison concerning at least one characteristic. 
On the other hand, those tools that did produce useful results show promising extraction correctness, especially when also considering the combinations of individual tools.
For all three evaluated metrics (precision, recall, and F1-measure), at least one tool or combination of tools exhibited a value of over 0.90, which we consider to be a generally good score without further context.

It is a known issue of static analysis security testing (SAST) tools, that they often produce a high number of false positive warnings, which is a major factor in inhibiting their adoption by practitioners~\cite{Christakis16_program_analysis_study,Johnson13_devs_use_sast}.
In the context of this work, falsely detected characteristics could likely have a comparable effect on the user-experience.
Some evaluated tools exhibited concerning behavior in this regard, for example, \textit{Attack Graph Generator} with more false FPs than TPs in the detection of connections and an overall precision of 0.48.
Nevertheless, there are four tools with an overall precision above 0.90 (\textit{Code2DFD}: 0.91, \textit{MicroDepGraph}: 0.99, \textit{microMiner}: 0.97, and \textit{RAD}: 0.98) and three combinations of tools (All\textsubscript{AND}: 0.97, Best\textsubscript{P}: 0.99, and Best\textsubscript{F1}: 0.93).

In some use-cases, a high recall might be more important than the precision. 
For this metric, the best-performing individual tool \textit{Attack Graph Generator} achieves a value of 0.86 over all characteristics in its extraction scope, which could be considered low depending on the scenario.
Here, combining multiple tools proves to be a valuable approach, with the combination Best\textsubscript{R} consisting of \textit{Attack Graph Generator}, \textit{Code2DFD}, \textit{RAD}, and \textit{RAD-source} exhibiting a recall of 0.96.

Finally, with the F1-score combining precision and recall into a single metric, it can be considered as a suited overall assessment criterion.
A value of 0.86 over all characteristics was observed for \textit{Code2DFD} and can be improved to a result of 0.91 when combining \textit{Code2DFD} with \textit{MicroDepGraph}, \textit{RAD}, and \textit{RAD-source}.
Whether this is seen as sufficiently correct has to be decided for a specific use-case, however, we see it as a reasonably good result, especially when considering that the tools show potential to be improved (see Section~\ref{sub:lessons_learned}, lesson learned 3 and 4).

Concerning the tools' execution times, all tools that produced meaningful results showed short execution times per application.
\textit{RAD}, \textit{RAD-source}, \textit{Prophet}, and \textit{microMiner} all average well below a second per application, \textit{Attack Graph Generator} and \textit{MicroDepGraph} around 1.5 seconds, and \textit{Code2DFD} averages 7.3 seconds per application.
Only \textit{MicroGraal} -- which also did not produce any results -- takes a full minute per application to run on average.
These execution times should not pose any problems for most scenarios.
The distribution of the execution times also resembles the complexity of the tools' analysis approaches.
\textit{Code2DFD} with the most in-depth approach takes the longest to run, while the tools simply parsing deployment files are much quicker.

Regarding the use of multiple tools in combination over a single tool, the observed results suggest that it is beneficial to do the former in most cases.
\textit{Code2DFD} is the only tool that showed extraction correctness that could not be improved by combining it with other tools, but only for the characteristic components and only in F1-score.
For the other characteristics and metrics, a combination was found that outperformed the best individual tool for this case.

We note that the reported execution times for combinations of tools are simply the sum of the individual tools' execution times. 
The numbers could show to be higher in reality because of added overhead due to the integration and, e.g., a voting mechanism, but that they could also be improved if the implementations are truly joint into a single tool combining the multiple analysis techniques.
With the observed times as the basis for an assessment, we can say that they should be sufficiently low for a use of the tools in most analysis settings. 

A look into the individual tools' performance -- especially those that failed to produce meaningful results -- shows,

\subsection{Lessons Learned}
\label{sub:lessons_learned}

Some general observations derive from the detailed discussion of the results above.
They are presented below as \textit{lessons learned} and could indicate where future research efforts might best be pointed.

\paragraph{\textbf{Lesson learned 1: Existing tools' reproducibility is limited.}}

Although we did not investigate the property of reproducibility systematically, the descriptions of our experience in executing the tools and the missing results for many tools despite them having executed successfully show issues in this regard.
While some tools could be run without hindrance, others required some adjustments or debugging effort or did not at all perform in the way they were presented in their corresponding publications.
Although most of the tools in our study are prototypes for academic purposes instead of carefully maintained products, this indicated degradation of reproducibility is concerning and an issue both in the academic sense and for the adoption of the approaches by practitioners.

\paragraph{\textbf{Lesson learned 2: Basic architecture extraction can be achieved with high precision and in short time via simple parsing of deployment files.}}

Parsing deployment files to extract the analyzed systems' basic architecture (i.e., components and/or connections) is the most common employed technique amongst the tools in our comparison.
All but one tool apply this technique: \textit{Attack Graph Generator}, \textit{Code2DFD}, and \textit{MicroDepGraph} parse Docker Compose files, \textit{Code2DFD} also parses Docker files, and \textit{MicroMiner} parses Kubernetes files.
In fact, for the comparison of tools, the results of the \textit{Attack Graph Generator} and \textit{microMiner} are solely achieved via this technique.
The high precision of \textit{microMiner} in extracting components in this way and the prevalence of the technique amongst the tools is a testament to its effectiveness.
Deployment files are simple, unambiguous documents that can be easily and very efficiently parsed, resulting in high precision and very short execution times of the tools relying on this technique.
Considering that only a few different containerization and orchestration solutions could cover a large part of microservice applications emphasizes the benefits of this technique.

\paragraph{\textbf{Lesson learned 3: Achieving a high recall requires a deeper analysis of the source code beyond deployment files and slightly longer execution times.}}

Although deployment files enable tools to achieve high precision with a very simple detection technique, they critically do not capture all relevant system characteristics. 
Not only are whole groups of system characteristics not always or never detectable via these files (components are the only group that is always included), but also are individual instances of a group not always present.
There are a total of 65 components that are not included in a deployment file in the used dataset.
Their existence has to be detected deeper in the source code.
An approach followed by multiple tools is the identification of Java annotations. 
They are easily detectable and simultaneously have a high information content concerning the architecture extraction.
Other system characteristics can only be detected, e.g., based on the existence of dependencies or plugins.
To this end, following heuristics can yield good results, for example by inferring the existence of system characteristics based on the detection of the required import statements.
The importance of a deeper code analysis can be seen in the observed results, where only tools that go beyond the deployment files in their analysis achieve high values of recall.
This also leads to longer execution times, however, all observed times are still reasonably low and should not pose a problem for the majority of scenarios.

\paragraph{\textbf{Lesson learned 4: The biggest causes of false positives are inaccurate heuristics and flaws in the tools' implementations.}}

An investigation of the observed false positives revealed, that there are two main causes for them. 
First, using heuristics to drive the architecture extraction works well in many cases but can also backfire.
For example, \textit{Prophet} detects components via the analyzed repository's structure and creates a component for every folder in it.
While many repositories containing microservice applications on GitHub are structured such that each individual microservice is in a folder of its own, this approach also leads to many false positives because not every folder contains the code of a microservice.
The second main cause are flaws in the implementation of the tools. 
These are false positives that are not based on the underlying approaches of the tools, but result from an improper implementation of the approach.
This was especially prevalent for \textit{Code2DFD} with its more complex extraction logic compared to the other tools.

\paragraph{\textbf{Lesson learned 5: Improving the extraction accuracy to almost-perfect results seems possible with existing techniques.}}

Despite showing shortcomings or complete failures of individual tools, the presented results are promising when looking at the group of tools in whole.
Tools or combination of tools are available in the literature that achieve very good results for all evaluated metrics (precision, recall, and F1-score).
Further, an examination of all undetected system characteristics showed, that none of the characteristics exhibits any property that would make it infeasible to detect with either of the evaluated tools' techniques.
We believe, that the tools' underlying approaches are well-suited to identify all characteristics in the used dataset while keeping the number of false positives low.
A more careful implementation of the prototypes is needed for this, as mentioned in lesson learned 3.
Specifically, we believe that an approach that features parsers for a small set of different deployment files as a foundation, and enhances it with a deeper source code analysis should be able to exceed the currently observed results.
The tool \textit{Code2DFD} already employs parsing of different deployment files and deep source code analysis.
A more mature implementation of it than the current prototype combined with the inclusion of detection techniques from \textit{RAD} and \textit{RAD-source} should be able to yield a very high score in extracting all system characteristics across all evaluated metrics, according to our qualitative assessment.

\section{Limitations}
\label{sec:limitations}

\subsection{Deviations from Registered Reports}
\label{sub:deviations_from_rr}

This study was performed in strict accordance to the methodology presented in the registered report where possible.
There were, however, two minor, unforeseen deviations necessary which were imposed by the analyzed tools.

First, we had expected that all identified tools would have components and connections in their extraction scopes, and that all could be compared on these two characteristics at least.
As shown in Table~\ref{tbl:characteristics_matrix}, only five of the nine tools consider components and eight consider connections in their architecture extraction.
Further, only four tools showed results that can be considered somewhat effective for the extraction of components and three tools for the extraction of connections.

Second, we did not extend the dataset with ground truth for all characteristics in the extraction scopes of the identified tools.
Specifically, we omitted this step for the characteristics CRUD operations (only considered by \textit{authz-flow-analysis}), attack graphs (only considered by \textit{Attack Graph Generator}), and access control inconsistencies (only considered by \textit{MicroGraal} and \textit{Prophet}).
There are multiple reasons for the exclusion: for characteristics that are only considered by a single tool, a comparison of extraction accuracy would not have been possible; the extraction of access control inconsistencies in the tools is not fully automated but follows a human-in-the-loop approach; and the tool considering CRUD operations had to be excluded from the comparison.

Finally, a new research question has been added compared to the registered report (RQ5, concerning the combinations of tools).
We believe it to be a valuable addition to the presented results, and there is no impact on any other parts of the study.
The methodology is a natural extension of the one that has been peer-reviewed, and the sections that were not part of the registered report are clearly marked.
Therefore, we see no introduced threats to validity or malpractice concerning the process of registered reports in this.

\subsection{Anticipated Risks}
\label{sub:risks}

In our registered report, we had anticipated three risks that could have occurred during the conduction of the planned study.
Now, after having executed all steps, we revisit these risks in the following.

\textit{\textbf{Risk 1: Inability to Execute Tools.}} While some tools posed difficulties in running them, we received valuable support from all authors we had contacted and were then successful in executing most of them.
However, one tool (\textit{Prophet2}) could not be considered further in the study due to our inability to run it.
A detailed description of the problem is given in Section~\ref{sub:identified_tools}. 
In short, the tool has been abandoned since its publication, we were not able to fix the tool on our own, and the authors have moved on from academia.

From those tools that were part of the comparison, two were not successful in extracting producing any results (\textit{MicroGraal} and \textit{ContextMap}) and others extracted only few individual characteristics from their extraction scope (\textit{Prophet}, \textit{RAD}, \textit{RAD-source}).
However, all tools indicated a successful execution despite producing no or very limited results and are thus seen as valid subjects.

\textit{\textbf{Risk 2: Tools' Extraction Scopes too Distinct for Comparison.}} For the case that the identified tools have extraction scopes that are distinct from each other and thus not allow any comparison of extracted characteristics, we had planned to compare them on the characteristics components and connections, as we believed these to be in the extraction scope of every architecture recovery tool.
The study showed, that this assumption is not true for all tools (see Table~\ref{tbl:characteristics_matrix}).
Some require the list of components to be provided, and one does not consider connections.
Nevertheless, the three evaluated characteristics are in the extraction scopes of five or more tools, thus providing a valuable comparison.

Across all identified tools, there are three additional system characteristics that are only extracted by a single tool or two tools each (CRUD operations by \textit{authz-flow-analysis}, which was excluded anyway; attack graphs by \textit{Attack Graph Generator}; and access control inconsistencies by \textit{MicroGraal} and \textit{Prophet}).
The anticipated risk hence did materialize, as the tools could not be compared concerning the extraction accuracy for these additional characteristics.

\textit{\textbf{Risk 3: Characteristics not existent in Dataset.}} 
The used dataset was extended with full information on endpoints in the contained applications, as described in Section~\ref{sub:extension_of_dataset}.
Three additional system characteristics are in the extraction scope of one or two of the identified tools, but not existent in the ground truth dataset.
Additionally, for two of them, it is unclear how the ground truth could be created manually.
We therefore decided not to extend the dataset with them, as no comparison would have been possible.

One other case regarding this anticipated risk is the tool \textit{protoc-gen-scip}, which was identified in the literature but excluded from the comparison.
The reason is described in Section~\ref{sub:identified_tools}. In summary, it is made for microservice applications using the RPC protocol.
The authors use one open-source application and one application made specifically for this use to evaluate their tool, and they mention the scarcity of applications using the RPC protocol.
Extending the dataset with new applications would have meant a considerable effort without a fair comparison of tools, since they would have been executed on different applications.
For this reason, this tool was not considered further in the study.

\subsection{Threats to Validity}
\label{sub:threats_to_validity}

The work in this paper is subject to some threats to validity that could have influenced the process and hence limit the presented results' validity.
They are presented in the following together with employed mitigations.
We recall that this paper's methodology has undergone peer-review before and was published as a registered report prior to the execution of the presented work.

\paragraph{\textbf{Internal Validity.}}

The authors' subjective judgement is needed at multiple places of a literature review.
The identification of tools via a multivocal literature review in this paper could therefore have been influenced by selection bias, search bias, and extraction errors during the selection and examination of sources.
We addressed these issues by following robust and established methods for performing literature reviews.
Specifically, all steps relying on subjective decisions were performed by two authors independently and conflicts solved with a third author, inter-rater agreement was measured and reported in terms of Cohen's Kappa, reproducible selection criteria and details on identifying tools in the selected sources were reported, and a replication package containing all data needed to evaluate the validity of the intermediary and final results is made available.
Despite these efforts, we might not have been successful in identifying all tools in the targeted scope of this work, e.g., because of publication bias which could prevent the publication of tools that follow an approach that is not sufficiently novel to be accepted in the academic literature.
The consideration of the gray literature in addition to academic sources is designed to mitigate this threat to validity, but could in turn be subject to database bias, i.e., could have been affected by the choice of gray literature sources that were analyzed.
The inclusion of broad sources (especially Google Search) was precisely intended to mitigate the above-described threat.

The quality of the gray literature sources was not evaluated or taken into consideration in the selection process, as is done in some gray literature reviews because these sources are not peer-reviewed per definition.
We decided against such a process because the existence of a reference to a tool that fits the scope of this study is not affected by the quality of the source.

The measurement of the tools' correctness in architecture recovery relied on manual work, both in the creation of the ground truth and in the analysis of the outputs.
However, the reliability of the used dataset's correctness is strengthened by the fact that the initial dataset has been published and peer-reviewed before, and that the extensions made in the context of this work are minor and were obtained by repeating activities in line with those of the original, peer-reviewed process.
The correctness of translating the tools' outputs into the chosen measurements and metrics is made more reliable by involving two authors in the process and by providing a replication package supporting all results.

Evaluating the tools' correctness was only possible for tools that could be executed successfully and that produced meaningful results.
One tool had to be excluded from the study based on our inability to run it, and two tools were executed and indicated a successful analysis but did not produce any results.
We spent considerable effort on these cases and sought advice from the authors, but ultimately did not succeed.
Although this may limit the expressiveness of the presented results, we believe that no further feasible actions could have been taken.

\paragraph{\textbf{External validity.}} 

The presented results and drawn conclusions might not entirely map to other execution scenarios.
All evaluated tools are either specific to Java or independent of the analyzed applications' programming language, which matches the applications in the used dataset.
Nevertheless, the dataset is the biggest factor that could have influenced the observed results' generalizability, since the applications are small- to medium-sized and show relatively homogeneous architectural patterns.
Thus, the tools could perform differently on bigger or more complex applications, and other conclusions could have been drawn concerning their comparison.
However, the dataset is the largest one currently available in the literature for this purpose.
Future work could include the creation of a dataset containing more industry-near applications, or a replication of the work if such a dataset is published by others.

Some identified tools extract more characteristics than the ones evaluated in the comparison, as reported.
Their full functionality for architecture recovery is hence not covered by the comparison, only the described characteristics are.

\paragraph{\textbf{Construct validity.}} 

The tools were evaluated based on precision, recall, and F1-measure, which are commonly used and objective metrics for this use.
They capture the tools' most important requirement, their correctness in architecture recovery.
Additionally, the tools' execution times were measured to provide a basis for assessing their suitability for being used in time-sensitive contexts.
Nevertheless, other metrics might be more important for specific use-cases, and the presented results could show a different performance when evaluated on these.

Most of the tools were initially presented as prototypes representing an analysis approach.
Therefore, the results in this paper can not be seen as evaluation of the approaches themselves, because the implementations might contain flaws and thus not accurately represent the intended approach.
Since this paper aims to provide a comparison of available tools and not approaches, this does not limit the validity of the drawn conclusions.
\section{Related Work}
\label{sec:related_work}

Various secondary studies have been presented in the realm of microservice architecture.
Most early work in this regard focused on delineating the similarities and differences between the microservice architecture and the monolithic architecture, service-oriented architecture, and other prior styles, as well as on outlining the trends, benefits, and downsides of the microservice architecture (e.g., Dragoni et al.~\cite{Dragoni16_microservices_yesterday_today_tomorrow} with their ``yesterday, today, and tomorrow'' of microservices or Soldani et al.~\cite{Soldani2018_MSAPainsGains} with their ``pains and gains'' of using the microservice architecture).

Other work considered the provided tool support of presented techniques early-on already.
In 2016 (two years after the emergence of the term ``microservice architecture'' following Lewis and Fowler \cite{Lewis14_microservices}), Pahl and Jamshidi conducted a systematic mapping study~\cite{pahl2016microservices} to review the existing literature on the microservice architecture at the time.
One of the considered aspects in their work was the tool support available for software engineering activities related to microservices.
The authors identified two tools for the migration of monolithic applications to the microservice architecture, but they note a general lack of tool support at the time of publishing their work.
Similarly, Alshuqarayn et al. \cite{alshuqayran2016systematic} conducted a systematic mapping study on the microservice architecture in the same year, identifying trends that were visible in the academic literature at the time. 
They identified a wide variety of modeling diagrams being used to represent the architecture of microservice applications, but no work on their recovery.

The usefulness of architecture recovery techniques for different use-cases have been noted by some authors.
For example, Abdelfattah and Cerny~\cite{Abdelfattah_2023} conducted a rapid review of the literature concerning reasoning in microservices -- a term covering assessment and understanding of an existing system, according to the authors -- and evolution of microservice systems.
They state architecture recovery to be at the core of the reasoning process and emphasize its necessity to obtain a holistic view of microservice applications.
Their description of analyzing the evolution of microservice applications is also based on a recovered architectural view.
Multiple approaches for architecture recovery are identified in the paper, some of which also provide implementations.
However, the existence of tool support is not a focus of the review.
Bushong et al.~\cite{Bushong_2021} performed a systematic mapping study on approaches for microservice analysis and architecture evolution.
Multiple of the techniques they identified have architecture reconstruction as their main goal and others as a means to achieve other goals such as analyzing security or system evolution.
One of the dimensions the authors used to classify the findings of their study is whether a tool is provided for proposed approaches.
Eleven tools were identified by the authors of the study across 55 presented papers.
One tool performs architecture reconstruction, however, dynamically (as was pointed out by Bakhtin et al.~\cite{Bakhtin23_tools_study}), and is therefore not in the scope of our study.

Most of the related work on microservice applications and their analysis present work on the same abstraction level as the architecture recovery tools targeted by our study create.
Although it is not explicitly mentioned in all cases, many of these findings arguably require an architecture recovery tool as a preliminary.
For example, both Neri et al.~\cite{Neri2020_MLRPrinciplesSmellsRefactoringsMSA} and Ponce et al.~\cite{Ponce2022_MLRSecuritySmellsMSA} conducted multivocal literature reviews on smells of microservice applications.
The former is focussed on architectural smells directly, while the latter is more security-oriented, and some of the identified security smells are at the architectural level as well.

The availability of tool support for identified approaches is often reported in literature reviews for other software engineering activities for microservice applications.
For example, Fritzsch et al.~\cite{Fritzsch19_refactoring_review} conducted a review of microservice refactoring approaches for the migration of applications from a monolithic to a microservice architecture.
The authors identified prototype implementations for three and a more mature tool for one of the ten compared approaches.
Saucedo et al.~\cite{saucedo2024migration} also targeted the migration of monoliths to microservices with a systematic mapping study, but are more generally concerned with depicting the typical activities of this process.
They identified 31 tools that are used in refactoring activities, including general utilities such as tools for source control.
Gortney et al.~\cite{Gortney22_visualizing_microservices} present the findings of a systematic mapping analysis of architecture visualization approaches that perform dynamic analysis.
From the 20 identified approaches, 13 are tool-supported.
A systematic literature review by Lelovic et al.~\cite{lelovic2024change} found 19 tools for change impact analysis in microservice applications.
Bakhtin et al.~\cite{Bakhtin22_api_patterns} performed a gray literature review of tools detecting API patterns, finding 59 tools.
A gray literature review presented by Giamatti et al.~\cite{GIAMATTEI2024111906} identified 71 tools for monitoring microservice applications.
Work closely related to our study was published by Cerny et al.~\cite{Cerny22_microservice_reconstruction_review}, who conducted a review of static and dynamic microservice architecture reconstruction approaches and visualization techniques.
Finally, the systematic mapping study by Bakhtin et al.~\cite{Bakhtin23_tools_study} that this paper is based on also identified tools for architecture recovery.
All these reviews, however, present the identified techniques and tools entirely based on information reported in the corresponding publications.
None of them executed the found tools and report on the observations, as we did in this paper.

Studies comparing tools based on results observed when executing them have been presented in the literature for other tasks than architecture recovery and other domains than microservices.
For example, this general study methodology has been used to compare static analysis tools for quality assurance by Mantere et al.~\cite{Mantere09_comparison_static_analysis_tools}, by Lenarduzzi et al.~\cite{LENARDUZZI2023111575}, by Li et al.~\cite{Li23_comparison_sast}, and by Liu et al.~\cite{Liu23_quality_assurance_tools} or to compare bug detection tools by Habib and Pradel~\cite{Habib18_bug_detection}, by Tomassi~\cite{Tomassi18_bug_detection}, and by Thung et al.~\cite{Thung15_detect_defects}.
Even comparisons of tools for software architecture recovery have been presented, e.g., by Lutellier et al.~\cite{Lutellier15_comparing_sar_techniques,Lutellier18_code_dependencies_sar} and by Garcia et al.~\cite{Garcia13_sar_comparison}.
Although the above studies base their findings on executing the compared tools, as we did as well, they concern different domains than the microservice architecture.

Finally, two recent publications also follow the approach of executing identified tools to compare them, and they both target the microservice architecture domain.
First, Akkaya and Ovatman~\cite{akkaya2022comparative} identified three tools for microservice decomposition -- i.e., tools for the migration of monolithic applications to the microservice architecture -- from multiple prior literature reviews and executed them on the same set of applications.
Then, they compared the results against a manual decomposition of the applications.
Second, Wang et al.~\cite{wang2024microservice} selected eight microservice decomposition tools from the results of a literature review and executed them on a set of applications.
The authors then compare the tools based on the observed results concerning several relevant metrics.
Although these studies target tools for a different use-case than we did in our work, both also identified the need for a comparison of tools based on observed results on a common dataset instead of reported results and specifications.

In conclusion, to the best of our knowledge, the presented work is the first review of static architecture recovery tools for microservice applications that provides a comprehensive comparison of the tools' performance on a common dataset.

\section{Conclusion}
\label{sec:conclusion}

This paper presented a comparison of static analysis architecture recovery tools for microservice applications based on their results observed when executing them on a common dataset.
A previous literature review by Bakhtin et al.~\cite{Bakhtin23_tools_study} has been repeated and extended into a multivocal literature review.
It resulted in the identification of 13 such architecture recovery tools, which have been characterized concerning the characteristics that they are intended to extract from analyzed applications (their \textit{extraction scope}).

When trying to run the tools, we faced various obstacles for many of them.
While no sound methodology for evaluating their usability was followed, our general experience in this step showed a concerning impression of their ease-of-use and the reproducibility of reported results.
Nine of the identified tools could successfully be run and were each executed on 17 microservice applications from the microSecEnD dataset~\cite{Schneider23_microsecend}.
The dataset contains dataflow diagrams which serve as ground truth for evaluating the tools' extraction accuracy for the characteristics components, connections, and endpoints.
The DFDs in the dataset have been extended with model items for endpoints for this purpose.
By comparing the tools' created outputs to the ground truth, we calculated precision, recall, and F1-score of each tool as measures of their correctness in architecture recovery.
For each of the three considered characteristics, the results of all tools that have it in their extraction scope were compared.
The best-performing tools in terms of precision, recall, and F1-measure over all characteristics in their extraction scope are \textit{MicroDepGraph}~\cite{Rahman19_microdepgraph} (P = 0.99), \textit{Attack Graph Generator}~\cite{Ibrahim19_attack-graph-generator} (R = 0.86), and \textit{Code2DFD}~\cite{Schneider23_code2dfd} (F1 = 0.86), respectively.

We also investigated whether it is possible to achieve a higher score in each metric by combining the results of multiple tools.
To this end, the results of different combinations of tools were merged with logical ``OR'' and ``AND'' operations on the individual characteristic level.
The results show, that for each metric and each characteristic, a tool combination can be found that outperforms or at least matches the accuracy of the best individual tool.
Specifically, a combination of \textit{Code2DFD}~\cite{Schneider23_code2dfd}, \textit{MicroDepGraph}~\cite{Rahman19_microdepgraph}, \textit{RAD}~\cite{Das21_rad}, and \textit{RAD-source}~\cite{Das21_rad} showed the highest F1-score (F1 = 0.91) of all evaluated combinations over all characteristics.

The presented work depicts the current state-of-the-art in the domain and can be a valuable resource for researchers and practitioners alike.
We summarize our insights and findings in five \textit{lessons learned}, which point out effective techniques and current obstacles, and give pointers for improving static analysis architecture recovery tools for microservice applications in future work.

\section*{Data Availability}
\noindent
A replication package of all relevant data is available at Zenodo~\cite{Replication_package} and GitHub~\footnote{https://github.com/M3SOulu/SARbenchmarks}.

\section*{Conflict of interest}
The authors declare that they have no conflict of interest.

\begin{acknowledgements}
We would like to express our gratitude to those authors we contacted for help in running their tools and who all answered supportively.
This material is based upon work supported by grants from the Research Council of Finland (grants n. 349487 and 349488 - MuFAno).
\end{acknowledgements}

\bibliographystyle{IEEEtran}
\bibliography{tool_comparison}

\begin{thebibliography}{10}
\providecommand{\url}[1]{#1}
\csname url@samestyle\endcsname
\providecommand{\newblock}{\relax}
\providecommand{\bibinfo}[2]{#2}
\providecommand{\BIBentrySTDinterwordspacing}{\spaceskip=0pt\relax}
\providecommand{\BIBentryALTinterwordstretchfactor}{4}
\providecommand{\BIBentryALTinterwordspacing}{\spaceskip=\fontdimen2\font plus
\BIBentryALTinterwordstretchfactor\fontdimen3\font minus \fontdimen4\font\relax}
\providecommand{\BIBforeignlanguage}[2]{{%
\expandafter\ifx\csname l@#1\endcsname\relax
\typeout{** WARNING: IEEEtran.bst: No hyphenation pattern has been}%
\typeout{** loaded for the language `#1'. Using the pattern for}%
\typeout{** the default language instead.}%
\else
\language=\csname l@#1\endcsname
\fi
#2}}
\providecommand{\BIBdecl}{\relax}
\BIBdecl

\bibitem{Tukaram22_security_rules}
A.~Bambhore~Tukaram, S.~Schneider, N.~E. D\'{\i}az~Ferreyra, G.~Simhandl, U.~Zdun, and R.~Scandariato, ``Towards a security benchmark for the architectural design of microservice applications,'' in \emph{ARES}.\hskip 1em plus 0.5em minus 0.4em\relax New York, NY, USA: ACM, 2022.

\bibitem{Ponce2022_MLRSecuritySmellsMSA}
F.~Ponce, J.~Soldani, H.~Astudillo, and A.~Brogi, ``Smells and refactorings for microservices security: A multivocal literature review,'' \emph{JSS}, 2022.

\bibitem{Taibi20_microservices_antipatterns}
D.~Taibi, V.~Lenarduzzi, and C.~Pahl, \emph{Microservices Anti-patterns: A Taxonomy}.\hskip 1em plus 0.5em minus 0.4em\relax Cham: Springer International Publishing, 2020.

\bibitem{Cao24_catma}
C.~Cao, S.~Schneider, N.~Diaz~Ferreyra, S.~Verweer, A.~Panichella, and R.~Scandariato, ``Catma: Conformance analysis tool for microservice applications,'' in \emph{ICSE-Companion}, 2024.

\bibitem{Dragoni16_microservices_yesterday_today_tomorrow}
N.~Dragoni, S.~Giallorenzo, A.~Lluch-Lafuente, M.~Mazzara, F.~Montesi, R.~Mustafin, and L.~Safina, \emph{Microservices: yesterday, today, and tomorrow}.\hskip 1em plus 0.5em minus 0.4em\relax Springer International Publishing, 2016.

\bibitem{Lewis14_microservices}
\BIBentryALTinterwordspacing
J.~Lewis and M.~Fowler, ``Microservices: a definition of this new architectural term,'' 2014. [Online]. Available: \url{https://martinfowler.com/articles/microservices.html}
\BIBentrySTDinterwordspacing

\bibitem{DiFrancesco19_architecting_microservices}
P.~{Di Francesco}, P.~Lago, and I.~Malavolta, ``Architecting with microservices: A systematic mapping study,'' \emph{JSS}, 2019.

\bibitem{Soldani2018_MSAPainsGains}
J.~Soldani, D.~A. Tamburri, and W.-J. {Van Den Heuvel}, ``The pains and gains of microservices: A systematic grey literature review,'' \emph{JSS}, 2018.

\bibitem{Arisholm06_impact_uml}
E.~Arisholm, L.~C. Briand, S.~E. Hove, and Y.~Labiche, ``The impact of uml documentation on software maintenance: an experimental evaluation,'' \emph{TSE}, 2006.

\bibitem{Budgen11_uml_slr}
D.~Budgen, A.~J. Burn, P.~Brereton, A.~B. Kitchenham, and R.~Pretorius, ``Empirical evidence about the uml: a systematic literature review,'' \emph{Software: Practice and Experience}, 2011.

\bibitem{Gravino10_empirical_investigation_code_comprehension}
C.~Gravino, G.~Tortora, and G.~Scanniello, ``An empirical investigation on the relation between analysis models and source code comprehension,'' in \emph{SAC}.\hskip 1em plus 0.5em minus 0.4em\relax ACM, 2010.

\bibitem{Gravino15_code_comprehension_uml}
C.~Gravino, G.~Scanniello, and G.~Tortora, ``Source-code comprehension tasks supported by uml design models: Results from a controlled experiment and a differentiated replication,'' \emph{Journal of Visual Languages \& Computing}, 2015.

\bibitem{Schneider24_DFDs_empirical_experiment}
S.~Schneider, N.~E. Diaz~Ferreyra, P.-J. Queval, G.~Simhandl, U.~Zdun, and R.~Scandariato, ``How dataflow diagrams impact software security analysis: an empirical experiment,'' in \emph{SANER}, 2024.

\bibitem{Alshuqayran18}
N.~Alshuqayran, N.~Ali, and R.~Evans, ``Towards micro service architecture recovery: An empirical study,'' in \emph{ICSA}, 2018.

\bibitem{Granchelli17}
G.~Granchelli, M.~Cardarelli, P.~Di~Francesco, I.~Malavolta, L.~Iovino, and A.~Di~Salle, ``Towards recovering the software architecture of microservice-based systems,'' in \emph{ICSAW}, 2017.

\bibitem{Kleehaus18}
M.~Kleehaus, {\"O}.~Uludag, P.~Sch{\"a}fer, and F.~Matthes, ``Microlyze: A framework for recovering the software architecture in microservice-based environments,'' in \emph{Information Systems in the Big Data Era}.\hskip 1em plus 0.5em minus 0.4em\relax Springer International Publishing, 2018.

\bibitem{Queval23_extracting_architecture}
P.-J. Qu{\'e}val and U.~Zdun, ``Extracting the architecture of microservices: An approach for explainability and traceability,'' in \emph{ECSA}.\hskip 1em plus 0.5em minus 0.4em\relax Cham: Springer Nature Switzerland, 2023.

\bibitem{Soldani21}
J.~Soldani, G.~Muntoni, D.~Neri, and A.~Brogi, ``The mtosca toolchain: Mining, analyzing, and refactoring microservice‐based architectures,'' \emph{Software: Practice and Experience}, 2021.

\bibitem{Schneider24_registered_report}
\BIBentryALTinterwordspacing
S.~Schneider, A.~Bakhtin, X.~Li, J.~Soldani, A.~Brogi, T.~Cerny, R.~Scandariato, and D.~Taibi, ``Comparison of static analysis architecture recovery tools for microservice applications,'' 2024. [Online]. Available: \url{https://arxiv.org/abs/2403.06941}
\BIBentrySTDinterwordspacing

\bibitem{Bakhtin23_tools_study}
A.~Bakhtin, X.~Li, J.~Soldani, A.~Brogi, T.~Cerny, and D.~Taibi, ``Tools reconstructing microservice architecture: A systematic mapping study,'' in \emph{Software Architecture. ECSA 2023 Tracks, Workshops, and Doctoral Symposium}, B.~Tekinerdo{\u{g}}an, R.~Spalazzese, H.~S{\"o}zer, S.~Bonfanti, and D.~Weyns, Eds.\hskip 1em plus 0.5em minus 0.4em\relax Cham: Springer Nature Switzerland, 2024, pp. 3--18.

\bibitem{GAROUSI2019101}
V.~Garousi, M.~Felderer, and M.~V. Mäntylä, ``Guidelines for including grey literature and conducting multivocal literature reviews in software engineering,'' \emph{IST}, 2019.

\bibitem{Moreschini2023TowardEM}
S.~Moreschini, G.~Recupito, V.~Lenarduzzi, F.~Palomba, D.~H{\"a}stbacka, and D.~Taibi, ``Toward end-to-end mlops tools map: A preliminary study based on a multivocal literature review,'' \emph{ArXiv}, 2023.

\bibitem{PELTONEN2021106571}
S.~Peltonen, L.~Mezzalira, and D.~Taibi, ``Motivations, benefits, and issues for adopting micro-frontends: A multivocal literature review,'' \emph{IST}, 2021.

\bibitem{Kitchenham04_procedures}
B.~Kitchenham, ``Procedures for performing systematic reviews,'' \emph{Keele, UK, Keele Univ.}, vol.~33, 08 2004.

\bibitem{Kitchenham07_guidelines}
B.~Kitchenham and S.~Charters, ``Guidelines for performing systematic literature reviews in software engineering,'' vol.~2, 01 2007.

\bibitem{Ralph20_empirical_standards}
P.~Ralph, N.~b. Ali, S.~Baltes, D.~Bianculli, J.~Diaz, Y.~Dittrich, N.~Ernst, M.~Felderer, R.~Feldt, A.~Filieri \emph{et~al.}, ``Empirical standards for software engineering research,'' \emph{arXiv preprint arXiv:2010.03525}, 2020.

\bibitem{10.1023/A:1009820201126}
K.~E. Emam, ``Benchmarking kappa: Interrater agreement in software processassessments,'' \emph{EMSE}, 1999.

\bibitem{Wohlin}
C.~Wohlin, ``Guidelines for snowballing in systematic literature studies and a replication in software engineering,'' in \emph{EASE}.\hskip 1em plus 0.5em minus 0.4em\relax ACM, 2014.

\bibitem{Schneider23_microsecend}
S.~Schneider, T.~\"Ozen, M.~Chen, and R.~Scandariato, ``microsecend: A dataset of security-enriched dataflow diagrams for microservice applications,'' in \emph{MSR}, 2023.

\bibitem{JetBrains22}
\BIBentryALTinterwordspacing
JetBrains, ``The state of developer ecosystem 2022,'' JetBrains, Tech. Rep., 2022, accessed on 09.02.2024. [Online]. Available: \url{https://www.jetbrains.com/lp/devecosystem-2022/microservices/}
\BIBentrySTDinterwordspacing

\bibitem{JRebel22}
\BIBentryALTinterwordspacing
JRebel, ``2022 java developer productivity report,'' JRebel, Tech. Rep., 2022, accessed on 09.02.2024. [Online]. Available: \url{https://www.jrebel.com/resources/java-developer-productivity-report-2022}
\BIBentrySTDinterwordspacing

\bibitem{Landis77_agreement}
J.~R. Landis and G.~G. Koch, ``The measurement of observer agreement for categorical data,'' \emph{Biometrics}, vol.~33, no.~1, pp. 159--174, 1977.

\bibitem{kozak2024micrograal}
R.~Hutcheson, A.~Blanchard, N.~Lambaria, J.~Hale, A.~E. David~Kozak, and T.~Cerny, ``Software architecture reconstruction for microservice systems using static analysis via graalvm native image,'' in \emph{SANER 2024}, ser. SANER.\hskip 1em plus 0.5em minus 0.4em\relax Institute of Electrical and Electronics Engineers, Mar. 2024.

\bibitem{abdelfattah2023towards}
A.~Abdelfattah, M.~Schiewe, J.~Curtis, T.~Cerny, and E.~Song, ``Towards security-aware microservices: On extracting endpoint data access operations to determine access rights,'' in \emph{13th international conference on cloud computing and services science (CLOSER 2023)}, 2023.

\bibitem{Ibrahim19_attack-graph-generator}
A.~Ibrahim, S.~Bozhinoski, and A.~Pretschner, ``Attack graph generation for microservice architecture,'' in \emph{Symposium on Applied Computing}.\hskip 1em plus 0.5em minus 0.4em\relax ACM, 2019.

\bibitem{Schneider23_code2dfd}
S.~Schneider and R.~Scandariato, ``Automatic extraction of security-rich dataflow diagrams for microservice applications written in java,'' \emph{JSS}, 2023.

\bibitem{Rahman19_microdepgraph}
M.~I. Rahman, S.~Panichella, and D.~Taibi, ``A curated dataset of microservices-based systems,'' 2019.

\bibitem{Muntoni21_microminer}
G.~Muntoni, J.~Soldani, and A.~Brogi, ``Mining the architecture of microservice-based applications from their kubernetes deployment,'' in \emph{Advances in Service-Oriented and Cloud Computing}.\hskip 1em plus 0.5em minus 0.4em\relax Cham: Springer International Publishing, 2021.

\bibitem{Soldani23_microtom}
J.~Soldani, J.~Khalili, and A.~Brogi, ``Offline mining of microservice-based architectures (extended version),'' \emph{SN Comput. Sci.}, 2023.

\bibitem{Bushong21_prophet}
V.~Bushong, D.~Das, A.~Al~Maruf, and T.~Cerny, ``Using static analysis to address microservice architecture reconstruction,'' in \emph{ASE}, 2021.

\bibitem{schiewe2022advancing}
M.~Schiewe, J.~Curtis, V.~Bushong, and T.~Cerny, ``Advancing static code analysis with language-agnostic component identification,'' \emph{IEEE Access}, vol.~10, pp. 30\,743--30\,761, 2022.

\bibitem{fang2023rpcover}
A.~Fang, R.~Zhou, X.~Tang, and P.~He, ``Rpcover: Recovering grpc dependency in multilingual projects,'' in \emph{2023 38th IEEE/ACM International Conference on Automated Software Engineering (ASE)}.\hskip 1em plus 0.5em minus 0.4em\relax IEEE, 2023, pp. 1930--1939.

\bibitem{Das21_rad}
D.~Das, A.~Walker, V.~Bushong, J.~Svacina, T.~Cerny, and V.~Matyas, ``On automated rbac assessment by constructing a centralized perspective for microservice mesh,'' \emph{PeerJ Computer Science}, vol.~7, 2021.

\bibitem{Christakis16_program_analysis_study}
M.~Christakis and C.~Bird, ``What developers want and need from program analysis: an empirical study,'' in \emph{Proceedings of the 31st IEEE/ACM International Conference on Automated Software Engineering}, ser. ASE '16.\hskip 1em plus 0.5em minus 0.4em\relax New York, NY, USA: Association for Computing Machinery, 2016, p. 332–343.

\bibitem{Johnson13_devs_use_sast}
B.~Johnson, Y.~Song, E.~Murphy-Hill, and R.~Bowdidge, ``Why don't software developers use static analysis tools to find bugs?'' in \emph{2013 35th International Conference on Software Engineering (ICSE)}, 2013, pp. 672--681.

\bibitem{pahl2016microservices}
C.~Pahl and P.~Jamshidi, ``Microservices: A systematic mapping study.'' \emph{CLOSER (1)}, pp. 137--146, 2016.

\bibitem{alshuqayran2016systematic}
N.~Alshuqayran, N.~Ali, and R.~Evans, ``A systematic mapping study in microservice architecture,'' in \emph{2016 IEEE 9th international conference on service-oriented computing and applications (SOCA)}.\hskip 1em plus 0.5em minus 0.4em\relax IEEE, 2016, pp. 44--51.

\bibitem{Abdelfattah_2023}
A.~S. Abdelfattah and T.~Cerny, ``Roadmap to reasoning in microservice systems: A rapid review,'' \emph{Applied Sciences}, vol.~13, no.~3, 2023.

\bibitem{Bushong_2021}
V.~Bushong, A.~S. Abdelfattah, A.~A. Maruf, D.~Das, A.~Lehman, E.~Jaroszewski, M.~Coffey, T.~Cerny, K.~Frajtak, P.~Tisnovsky, and M.~Bures, ``On microservice analysis and architecture evolution: A systematic mapping study,'' \emph{Applied Sciences}, 2021.

\bibitem{Neri2020_MLRPrinciplesSmellsRefactoringsMSA}
D.~Neri, J.~Soldani, O.~Zimmermann, and A.~Brogi, ``Design principles, architectural smells and refactorings for microservices: a multivocal review,'' \emph{SICS}, 2020.

\bibitem{Fritzsch19_refactoring_review}
J.~Fritzsch, J.~Bogner, A.~Zimmermann, and S.~Wagner, ``From monolith to microservices: A classification of refactoring approaches,'' in \emph{Software Engineering Aspects of Continuous Development and New Paradigms of Software Production and Deployment}.\hskip 1em plus 0.5em minus 0.4em\relax Cham: Springer International Publishing, 2019.

\bibitem{saucedo2024migration}
A.~M. Saucedo, G.~Rodr{\'\i}guez, F.~G. Rocha, and R.~P. dos Santos, ``Migration of monolithic systems to microservices: A systematic mapping study,'' \emph{Information and Software Technology}, p. 107590, 2024.

\bibitem{Gortney22_visualizing_microservices}
M.~E. Gortney, P.~E. Harris, T.~Cerny, A.~A. Maruf, M.~Bures, D.~Taibi, and P.~Tisnovsky, ``Visualizing microservice architecture in the dynamic perspective: A systematic mapping study,'' \emph{IEEE Access}, 2022.

\bibitem{lelovic2024change}
L.~Lelovic, A.~Huzinga, G.~Goulis, A.~Kaur, R.~Boone, U.~Muzrapov, A.~S. Abdelfattah, and T.~Cerny, ``Change impact analysis in microservice systems: A systematic literature review,'' \emph{Journal of Systems and Software}, p. 112241, 2024.

\bibitem{Bakhtin22_api_patterns}
A.~Bakhtin, A.~Al~Maruf, T.~Cerny, and D.~Taibi, ``Survey on tools and techniques detecting microservice api patterns,'' in \emph{SCC}, 2022.

\bibitem{GIAMATTEI2024111906}
L.~Giamattei, A.~Guerriero, R.~Pietrantuono, S.~Russo, I.~Malavolta, T.~Islam, M.~Dînga, A.~Koziolek, S.~Singh, M.~Armbruster, J.~Gutierrez-Martinez, S.~Caro-Alvaro, D.~Rodriguez, S.~Weber, J.~Henss, E.~F. Vogelin, and F.~S. Panojo, ``Monitoring tools for devops and microservices: A systematic grey literature review,'' \emph{JSS}, 2024.

\bibitem{Cerny22_microservice_reconstruction_review}
T.~Cerny, A.~S. Abdelfattah, V.~Bushong, A.~Al~Maruf, and D.~Taibi, ``Microservice architecture reconstruction and visualization techniques: A review,'' in \emph{SOSE}, 2022.

\bibitem{Mantere09_comparison_static_analysis_tools}
M.~Mantere, I.~Uusitalo, and J.~Roning, ``Comparison of static code analysis tools,'' in \emph{SECURWARE}, 2009.

\bibitem{LENARDUZZI2023111575}
V.~Lenarduzzi, F.~Pecorelli, N.~Saarimaki, S.~Lujan, and F.~Palomba, ``A critical comparison on six static analysis tools: Detection, agreement, and precision,'' \emph{JSS}, 2023.

\bibitem{Li23_comparison_sast}
K.~Li, S.~Chen, L.~Fan, R.~Feng, H.~Liu, C.~Liu, Y.~Liu, and Y.~Chen, ``Comparison and evaluation on static application security testing (sast) tools for java,'' in \emph{Proceedings of the 31st ACM Joint European Software Engineering Conference and Symposium on the Foundations of Software Engineering}, ser. ESEC/FSE 2023.\hskip 1em plus 0.5em minus 0.4em\relax New York, NY, USA: Association for Computing Machinery, 2023, p. 921–933.

\bibitem{Liu23_quality_assurance_tools}
H.~Liu, S.~Chen, R.~Feng, C.~Liu, K.~Li, Z.~Xu, L.~Nie, Y.~Liu, and Y.~Chen, ``A comprehensive study on quality assurance tools for java,'' in \emph{Proceedings of the 32nd ACM SIGSOFT International Symposium on Software Testing and Analysis}, ser. ISSTA 2023.\hskip 1em plus 0.5em minus 0.4em\relax New York, NY, USA: Association for Computing Machinery, 2023, p. 285–297.

\bibitem{Habib18_bug_detection}
A.~Habib and M.~Pradel, ``How many of all bugs do we find? a study of static bug detectors,'' in \emph{Proceedings of the 33rd ACM/IEEE International Conference on Automated Software Engineering}, ser. ASE '18.\hskip 1em plus 0.5em minus 0.4em\relax New York, NY, USA: Association for Computing Machinery, 2018, p. 317–328.

\bibitem{Tomassi18_bug_detection}
D.~A. Tomassi, ``Bugs in the wild: examining the effectiveness of static analyzers at finding real-world bugs,'' in \emph{Proceedings of the 2018 26th ACM Joint Meeting on European Software Engineering Conference and Symposium on the Foundations of Software Engineering}, ser. ESEC/FSE 2018.\hskip 1em plus 0.5em minus 0.4em\relax New York, NY, USA: Association for Computing Machinery, 2018, p. 980–982.

\bibitem{Thung15_detect_defects}
\BIBentryALTinterwordspacing
F.~Thung, Lucia, D.~Lo, L.~Jiang, F.~Rahman, and P.~T. Devanbu, ``To what extent could we detect field defects? an extended empirical study of false negatives in static bug-finding tools,'' \emph{Automated Software Engineering}, vol.~22, no.~4, pp. 561--602, 2015. [Online]. Available: \url{https://doi.org/10.1007/s10515-014-0169-8}
\BIBentrySTDinterwordspacing

\bibitem{Lutellier15_comparing_sar_techniques}
T.~Lutellier, D.~Chollak, J.~Garcia, L.~Tan, D.~Rayside, N.~Medvidovic, and R.~Kroeger, ``Comparing software architecture recovery techniques using accurate dependencies,'' in \emph{2015 IEEE/ACM 37th IEEE International Conference on Software Engineering}, vol.~2, 2015, pp. 69--78.

\bibitem{Lutellier18_code_dependencies_sar}
T.~Lutellier, D.~Chollak, J.~Garcia, L.~Tan, D.~Rayside, N.~Medvidović, and R.~Kroeger, ``Measuring the impact of code dependencies on software architecture recovery techniques,'' \emph{IEEE Transactions on Software Engineering}, vol.~44, no.~2, pp. 159--181, 2018.

\bibitem{Garcia13_sar_comparison}
J.~Garcia, I.~Ivkovic, and N.~Medvidovic, ``A comparative analysis of software architecture recovery techniques,'' in \emph{2013 28th IEEE/ACM International Conference on Automated Software Engineering (ASE)}, 2013, pp. 486--496.

\bibitem{akkaya2022comparative}
K.~Akkaya and T.~Ovatman, ``A comparative study of meta-data-based microservice extraction tools,'' \emph{IJSSMET}, 2022.

\bibitem{wang2024microservice}
Y.~Wang, S.~Bornais, and J.~Rubin, ``Microservice decomposition techniques: An independent tool comparison,'' in \emph{Proceedings of the 39th IEEE/ACM International Conference on Automated Software Engineering}, 2024, pp. 1295--1307.

\bibitem{Replication_package}
\BIBentryALTinterwordspacing
\emph{Replication package of the presented work}.\hskip 1em plus 0.5em minus 0.4em\relax Zenodo, 2024. [Online]. Available: \url{https://doi.org/10.5281/zenodo.14179613}
\BIBentrySTDinterwordspacing

\end{thebibliography}

\end{document}